\documentclass[11pt,american,twoside,a4paper]{article}
\usepackage[T1]{fontenc}
\usepackage[utf8]{inputenc}
\usepackage{latexsym}
\usepackage{amsmath}
\usepackage{bbm}
\usepackage{amssymb}
\usepackage{babel}
\usepackage{amsfonts}
\usepackage{fourier}
\usepackage{theorem}
\usepackage{epsfig}
\usepackage{enumerate}
\usepackage{color,xcolor}
\usepackage{xspace,needspace,xparse,easyReview}
\usepackage[active]{srcltx}
\usepackage[colorlinks=true]{hyperref}
\usepackage{fancyhdr}
\usepackage{xargs}
\usepackage[toc]{appendix}

\newcommandx{\ca}[2][1=]{\todo[inline,author={ca},
	linecolor=green,backgroundcolor=green!15,bordercolor=green,#1]{#2}}
\newcommandx{\canot}[2][1=]{\todo[author={ca},
	linecolor=green,backgroundcolor=green!15,bordercolor=green,#1]{#2}}
\colorlet{darkblue}{blue!50!black}
\hypersetup{
	colorlinks,%
	citecolor=darkblue,%
	filecolor=red,%
	linkcolor=darkblue,%
	urlcolor=darkblue,%
	pdfnewwindow=true,%
	pdfstartview={FitH}
}
\usepackage[style=cap-alpha,texencoding=UTF-8]{biblatex}
\numberwithin{equation}{section}

\newcounter{smallarabics}

\newcounter{smallroman}

\newcommand{\ben}{\begin{enumerate}[{\rm (1)}]}
\newcommand{\een}{\end{enumerate}}

\newtheorem{theorem}{Theorem}[section]
\newtheorem{proposition}[theorem]{Proposition}
\newtheorem{lemma}[theorem]{Lemma}
\newtheorem{definition}[theorem]{Definition}

\usepackage[normalem]{ulem}

\NewDocumentCommand{\ASS}{mm}{\expandafter\newcommand\csname #1\endcsname{{\hyperref[#1]{\bf (#2)}}}}
\NewDocumentCommand{\preASS}{mm}{\expandafter\newcommand\csname pre#1\endcsname{{\hyperref[#1]{\bf (#2)}}}}
\newcommand{\assuming}[3]{
\begin{quote}\label{#2}{\bf(#1) }%
#3%
\end{quote}%
\ASS{#2}{#1}}

\AtBeginEnvironment{theorem}{\Needspace{5\baselineskip}}
\AtBeginEnvironment{proposition}{\Needspace{3\baselineskip}}
\AtBeginEnvironment{definition}{\Needspace{5\baselineskip}}
\AtBeginEnvironment{corollary}{\Needspace{5\baselineskip}}
\AtBeginEnvironment{lemma}{\Needspace{5\baselineskip}}
\AtBeginEnvironment{quote}{\Needspace{5\baselineskip}}

\def\cA{{\mathcal A}} \def\cB{{\mathcal B}} 
 \def\cE{{\mathcal E}} 
 \def\cH{{\mathcal H}} 
 \def\cK{{\mathcal K}} \def\cL{{\mathcal L}}
 \def\cN{{\mathcal N}} \def\cO{{\mathcal O}}
\def\cP{{\mathcal P}}  \def\cR{{\mathcal R}}
\def\cS{{\mathcal S}}  
  \def\cX{{\mathcal X}}

\def\II{{\mathbb I}}
\def\RR{{\mathbb R}}
\def\ZZ{{\mathbb Z}}
\def\CC{{\mathbb C}}
\def\NN{{\mathbb N}}

\def\sS{\mathsf{S}}
\def\sR{\mathsf{R}}

\def\slim{\mathop{\mathrm{s-lim}}}

\def\shalf{{\scriptscriptstyle1/2}}

\def\tr{\mathop{\mathrm{tr}}\nolimits}
\def\sp{\mathop{\mathrm{sp}}}
\def\e{\mathrm{e}}
\def\i{\mathrm{i}}

\def\d{\mathrm{d}}

\def\bep{\begin{proposition}}
\def\eep{\end{proposition}}
\def\bet{\begin{theoreme}}
\def\eet{\end{theoreme}}
\def\bel{\begin{lemma}}
\def\eel{\end{lemma}}
\newcommand{\bra}{\langle}
\newcommand{\ket}{\rangle}
\newcommand{\ds}{\displaystyle}
\def\Ran{\mathop{\mathrm{Ran}}}
\def\Re{\mathop{\mathrm{Re}}}
\def\Im{\mathop{\mathrm{Im}}}
\def\Dom{\mathop{\mathrm{Dom}}}

\def\interior{\mathop{\mathrm{int}}}
\def\closure{\mathop{\mathrm{cl}}}
\newcommand{\one}{\mathbbm{1}}
\newcommand{\ttm}{\mathrm{2tm}}
\newcommand{\anc}{\mathrm{ancilla}}
\newcommand{\qpsc}{\mathrm{qpsc}}
\newcommand{\liouv}{\mathrm{Liouv}}
\newcommand{\koop}{\mathrm{Koop}}
\newcommand{\Id}{\mathrm{Id}}
\newcommand{\Mat}{\mathrm{Mat}}

\def\proof{\noindent {\bf Proof.}\ \ }
\def\qed{\hfill $\Box$\medip}

\def\textsl{{}}

\def\c0inf{C_0^\infty}
\def\proof{\noindent {\bf Proof.}\ \ }

\def\i{\mathrm{i}}
\newcommand{\beq}{\begin{equation}}
\newcommand{\eeq}{\end{equation}}
\newcommand{\bear}[1]{\begin{array}{#1}}
\newcommand{\ear}{\end{array}}

\renewcommand{\d}{\mathrm{d}}

\def\qed{$\Box$\medskip}

\def\bel{\begin{lemma}}
\def\eel{\end{lemma}}
\def\bet{\begin{theoreme}}
\def\eet{\end{theoreme}}
\def\bed{\begin{definition}}
\def\eed{\end{definition}}
\def\bep{\begin{proposition}}
\def\eep{\end{proposition}}

\def\bar{\overline}

\def\sp{{\rm sp}}
\def\Ent{{\rm Ent}}

\def\fr{{\mathfrak r}}

\def\fH{{\mathfrak H}}
\def\fM{{\mathfrak M}}
\def\fF{{\mathfrak F}}
\def\fS{{\mathfrak S}}
\def\bF{\boldsymbol{F}}

\def\dim{{\rm dim}}

\def\free{{\rm fr}}

\begin{document}
\def\today{}
\title{Entropic Fluctuations in Statistical Mechanics II. \\ Quantum Dynamical Systems}
\author{T. Benoist$^{1}$, L. Bruneau$^{2}$, V. Jak\v{s}i\'c$^{3}$, A. Panati$^4$, C.-A. Pillet$^4$
\\ \\
$^1$ Institut de Mathématiques de Toulouse, UMR5219,\\
 Université de Toulouse, CNRS, UPS, F-31062 Toulouse Cedex 9, France
\\ \\
$^2$ D\'epartement de Math\'ematiques, Universit\'e de Cergy-Pontoise\\
CNRS UMR 8088\\
95000 Cergy-Pontoise, France
\\ \\
$^3$Dipartimento di Matematica\\
Politecnico di Milano\\
piazza Leonardo da Vinci, 32 \\
20133 Milano,  Italy
\\ \\
$^4$Universit\'e de Toulon, CNRS, CPT, UMR 7332, 83957 La Garde, France\\
Aix-Marseille Univ, CNRS, CPT, UMR 7332, Case 907, 13288 Marseille, France
}
\maketitle
\thispagestyle{empty}
\noindent{\small{\bf Abstract.} The celebrated Evans--Searles, respectively
Gallavotti--Cohen, fluctuation theorem concerns certain universal statistical features of the
entropy production rate of a classical system in a transient, respectively
steady, state. In this paper, we consider and compare
several possible extensions of these fluctuation theorems to quantum systems. In
addition to the direct two-time measurement approach whose discussion is based
on \href{https://doi.org/10.1007/s11005-024-01777-0}{[LMP 114:32 (2024)]},
we discuss a variant where measurements are performed indirectly on an auxiliary
system called ancilla, and which allows to retrieve non-trivial statistical
information using ancilla state tomography. We also show that modular theory
provides a way to extend the classical notion of phase space contraction rate to
the quantum domain, which leads to a third extension of the fluctuation
theorems. We further discuss the quantum version of the principle of regular
entropic fluctuations, introduced in the classical context in
\href{http://dx.doi.org/10.1088/0951-7715/24/3/003}{[Nonlinearity 24, 699
(2011)]}. Finally, we relate the statistical properties of these
various notions of entropy production to spectral resonances of quantum
transfer operators. The obtained results shed a new light on the nature of
entropic fluctuations in quantum statistical mechanics.}

\newpage
\tableofcontents

\section{Introduction}
\label{sec-introduction-n}
This paper is a sequel to~\cite{Jaksic2011}, which was centered on two
celebrated results of classical non-equilibrium statistical mechanics: the
Evans--Searles and Gallavotti--Cohen {\sl Fluctuation Theorems}
\cite{Evans1994,Gallavotti1995b,Gallavotti1995c}. The main contribution
of~\cite{Jaksic2011} was to realize that these two physically and mathematically
distinct results were intimately related by a so-called {\sl Principle of
Regular Entropic Fluctuations} (abbreviated PREF in the following). The main
subject of the present work is the extension of these two fluctuation theorems
to the quantum domain, and a discussion of the PREF in this context.

{}From the perspective of classical statistical mechanics, both fluctuation
theorems involve a {\sl Large Deviation Principle} (LDP) for the time-averaged entropy
production observable ({\sl i.e.,} the phase space contraction rate) in the
large time limit. In the Evans--Searles case, the statistics is induced by a
reference, initial, state of the system\footnote{A state of a classical system
is a Borel probability measure on its phase space.} that evolves towards a
non-equilibrium steady state (NESS) in the  large time limit. The Evans--Searles
fluctuation theorem deals with a transient process, and for this reason is often
called {\sl Transient Fluctuation Theorem.} In the Gallavotti--Cohen case, the
statistics is directly induced by the NESS, and thus pertains to a stationary
process. Except in thermodynamically trivial situations, the reference state and
the NESS are mutually singular measures, and hence both physically and
mathematically the two fluctuation theorems are very different statements.

The starting point of the PREF is that, in spite of this difference,
in all known non-trivial examples where both theorems hold the respective LDP
rate functions are identical, and that this identity is equivalent to an
exchange of limits in the derivation of the LDP. The justification of this
exchange of limits is typically a deep dynamical problem whose validity was
raised in~\cite{Jaksic2011} to a principle: the PREF.

The classical fluctuation theorems come with equally celebrated {\sl Fluctuation
Relations.} If the system is {\sl Time-Reversal Invariant} (abbreviated TRI), the
Evans--Searles fluctuation relation asserts that the rate function $\II$
governing the large deviations of entropy production in the reference state
satisfies
\beq
\II(-s)=\II(s) +s
\label{es-r-in}
\eeq
on its domain. In the non-equilibrium steady state, according to the
Gallavotti--Cohen fluctuation relation, these large deviations are described by
a rate function $\II_+$ satisfying the same relation
\beq
\II_+(-s)=\II_+(s) +s
\label{gc-r-in}
\eeq
on its domain. The general mechanism behind~\eqref{es-r-in} is simple and this
relation is an immediate consequence of the LDP and time-reversal invariance;
see~\cite[Proposition 1.4]{Cuneo2017}, Proposition~\ref{es-thm-1} below, and the
comment after Theorem~\ref{thm-qspc-ti-cl-1}. In contrast, the only known
general mechanism behind~\eqref{gc-r-in} is the PREF that gives the equality
$\II=\II_+$, and so~\eqref{gc-r-in} is forced by~\eqref{es-r-in}. For further
discussion of these points we refer the reader to~\cite{Jaksic2011}.

Quantum fluctuation relations first appeared in the
works~\cite{Kurchan2000,Tasaki2000,Tasaki2003}. The LDP aspect was not
discussed in these works, and fluctuation relations were considered only for
finite times (see Theorem~\ref{gen-ttm-quant}(4) below). In~\cite{Kurchan2000,
Tasaki2000}, they were formulated in the context of finite quantum
systems\footnote{By {\sl finite quantum system} we mean a system whose
observables are linear operators acting on a finite-dimensional Hilbert space.}
and the main object of interest was the distribution of a random variable
expressing the change of entropy (or entropy production) in a two-time
measurement protocol. We will refer to this random variable as the {\sl
Two-Times Measurement Entropy Production} (abbreviated 2TMEP).
In~\cite{Tasaki2003}, the proposed fluctuation relation was formulated in the
general algebraic framework of quantum dynamical systems, and was phrased in
terms of the spectral measure of a suitable relative modular operator that
provided a non-commutative extension of the classical phase contraction along
the state trajectory. It turned out that these two proposals are identical and
that they provide a natural basis for a quantum Evans--Searles fluctuation
theorem and relation; see~\cite{Jaksic2010b} for a pedagogical introduction to
this topic and~\cite{DeRoeck2009} for an early work on the subject. The quantum
Evans--Searles fluctuation theorem and relation will be also briefly reviewed in
Section~\ref{sec-quantum-pref} below.

For a long time, a starting point for a quantum Gallavotti--Cohen fluctuation
theorem and quantum PREF in the same algebraic setting of quantum dynamical
systems was missing. The two main obstacles were:
\begin{enumerate}[(a)]
\item The limiting procedure that would allow to define the 2TMEP with respect
to NESS was unknown.
\item Any potential resolution of~(a) faces, in a more severe way, a problem
already present in the Evans--Searles case: the 2TMEP of large quantum
systems is not, even in principle, directly experimentally accessible.
\end{enumerate}

The point~(a) was recently resolved in~\cite{Benoist2023a,Benoist2024b}, and we
will quickly review the respective results in Section~\ref{sec-start-te-1}.
These works, however, are not concerned with the quantum fluctuation theorems.
In the present paper we build on~\cite{Benoist2023a}, showing that  the emerging
quantum Gallavotti--Cohen fluctuation theorem comes with an unexpected degree of
rigidity that makes it an immediate consequence of the quantum Evans--Searles
fluctuation theorem. The same applies to the equality of the respective rate
functions. This purely quantum phenomenon, due to the dominating decoherence
effect of the first measurement, essentially trivializes the resulting quantum
PREF.

Our second concern in this paper is point~(b). We resolve it by introducing
{\sl Entropic Ancilla State Tomography} (abbreviated EAST) which is, in principle,
experimentally accessible. This resolves the observability problem of the 2TMEP
with respect to {\sl any state}, including the NESS, and at the same time allows
for the introduction of a non-trivial quantum PREF that parallels the classical
one in its nature. This comes at the cost of a definition that does not ensure
an interpretation in terms of an entropy production random variable.

Our first set of results concerns the introduction of EAST for an
arbitrary initial state, and the associated quantum PREF (which we will call
{\sl strong\/}). In a second set of results, we characterize this
strong quantum PREF in terms of spectral resonances of a quantum transfer
operator. Here, we present an axiomatic approach to these resonances.
In the follow-up
work~\cite{Benoist2024c}, building on the techniques introduced
in~\cite{Jaksic1996b,Jaksic1996a,Jaksic2002a}, we illustrate this spectral
theory of the strong PREF on the example of the Spin--Fermion model which is one
of the paradigmatic models of open quantum systems (see the seminal
works~\cite{Davies1974,Spohn1978b}).

To our knowledge, the introduction of EAST in the context of quantum
fluctuations relations and quantum PREF is new and remains to be fully explored.
The results we have obtained about strong quantum PREF are surprising,
and we hope that they will trigger further investigations.
Indirect measurements have played and important role in recent experimental and
theoretical studies of quantum mechanical probabilistic rules, see
Section~\ref{sec-start-te-2X}. EAST falls under the umbrella of this direction
of research and its full theoretical and experimental significance needs to be
better understood.

We finish this introduction with three  additional remarks.
\begin{itemize}
\item The history of the subject is quite intricate, and some important
facts have being stated in the literature without proof. We have taken the
opportunity to close these holes, thus providing a reasonably self-contained
presentation of a complex topic. We refer the reader to
Remark 4 in Section~\ref{sec-remarks-tuluz} for further explanation and to
specific results.
\item The presentation we have chosen does not follow the most general possible
route, and we have attempted to strike a balance between emphasizing the
mathematical structure and focusing on physically relevant settings like open
quantum systems. The results formulated and proven in the specific setting of
open quantum systems can be easily generalized. We leave these generalizations
to the interested readers.
\item The main technical tool of this work is Araki's relative modular operator
theory~\cite{Araki1975/76,Araki1977,Araki1982}, which is perfectly suited for
non-commutative/quantum extensions of entropic notions in classical dynamical
systems and probability theory. This theory is discussed and reviewed in many
places in the literature\footnote{See~\cite{Jaksic2010b} for a pedagogical
introduction to this topic.}, and for definiteness we adopt, as we did in~\cite{
Benoist2023a}, the notation and conventions of~\cite[Section 6]{Jaksic2012}.
\end{itemize}

The paper is organized as follows. In
Sections~\ref{sec:repeat}--\ref{sec:repeat-2}, we briefly review the notation
and definitions of~\cite{Benoist2023a} which will be reused in the sequel. For
reason of space we can not enter into too many details, and we
strongly encourage the reader to consult this reference.
In Section~\ref{sec-start-te-1} we
introduce the two-time measurement entropy production and review the main
result of~\cite{Benoist2023a}. In Section~\ref{sec-start-te-2X} we introduce the
entropic ancilla state tomography. The quantum Evans--Searles and
Gallavotti--Cohen fluctuation theorems are discussed in
Section~\ref{sec-quantum-pref}, where we also introduce the quantum (weak and
strong) principle of regular entropic fluctuations. The classical theory of
entropic fluctuations based on the entropy production observable/phase space
contraction rate is reviewed in Section~\ref{sec-classical-1}. The quantum
entropy production observable/phase space contraction rate is discussed in
Section~\ref{sec-qpsc-1}. The classical and quantum case are compared in
Section~\ref{sec-cl-q-comp}. The quantum transfer operators are introduced in
Section~\ref{sec-quantum-tra}. The associated spectral resonance theory of the
strong quantum PREF is described in Section~\ref{sec-spectral-PREF}. The proofs
are given in Section~\ref{sec-proofs-sr}.

\smallskip
\paragraph*{Acknowledgments} The work of CAP and VJ  was partly  funded by  the
CY Initiative grant {\sl Investissements d'Avenir}, grant number ANR-16-IDEX-0008.
The work of TB was funded by the ANR project  {\sl ESQuisses}, grant number
ANR-20-CE47-0014-01, and  by the ANR project  {\sl Quantum Trajectories}, grant
number ANR-20-CE40-0024-01. VJ acknowledges the support of NSERC and  the support of the MUR grant "Dipartimento di Eccellenza 2023-2027" of Dipartimento di Matematica, Politecnico di Milano. A part of this
work was done during long term visits of LB and AP  to McGill and CRM-CNRS
International Research Laboratory IRL 3457 at University  of Montreal. The LB
visit was funded by the CNRS and AP visits by the CRM Simons and
FRQNT-CRM-CNRS programs. We also acknowledge the support of
the ANR project {\sl DYNACQUS}, grant number ANR-24-CE40-5714.

\section{Quantum dynamical systems}

\subsection{$C^\ast$-dynamical systems and their modular structure}
\label{sec:repeat}

In this and the following section we briefly review the mathematical description of
quantum dynamical systems that will be used in the present work, and describe the structure
of thermally driven open quantum systems which will serve as our paradigmatic examples.
We follow the conventions and notations of~\cite{Benoist2023a}
and refer the reader to this paper for further details and references.

\medskip
A $C^\ast\!${\sl-quantum dynamical system} is a triple $(\cO,\tau,\omega)$ where:
\begin{itemize}
\item $\cO$ is a $C^\ast$-algebra with a unit $\one$. Its elements $A\in\cO$
describe the observables of the system.
\item $\RR\ni t\mapsto\tau^t$ is a $C^\ast$-dynamics, {\sl i.e.,} a strongly
continuous group of $\ast$-automorphisms of $\cO$. It  describes the Heisenberg
time-evolution  $A_t=\tau^t(A)$ of the system observables. The infinitesimal
generator of a $C^\ast$-dynamics $\tau$  is a possibly unbounded
$\ast$-derivation $\delta$ of $\cO$. We use the convention
$\tau^t=\e^{t\delta}$.
\item $\omega$ is a state on $\cO$, {\sl i.e.,} an element of the closed convex
subset $\cS_\cO$ of the dual space $\cO^\ast$ consisting of linear functionals
$\omega\in\cO^\ast$ such that $\omega(A^\ast A)\ge0$ for all $A\in\cO$, and
$\omega(\one)=1$. We shall always equip $\cS_\cO$ with the weak$^\ast$-topology.
The number $\omega(A)$ is the quantum expectation value of the observable
$A\in\cO$, when the system is in the state $\omega$. A state $\omega\in\cS_\cO$
is faithful whenever $\omega(A^\ast A)=0$ implies $A=0$. States evolve according
to the Schrödinger picture $\omega_t=\omega\circ\tau^t$, so that
$\omega_t(A)=\omega(A_t)$. $\omega$ is called $\tau$-invariant
whenever $\omega_t=\omega$ for all $t\in\RR$.
\item Thermal equilibrium of the system at inverse temperature
$\beta\in\RR^\ast$ is described by a state $\omega$ satisfying the
$(\tau,\beta)$-KMS boundary condition: for any
$A,B\in\cO$ the function $F_{A,B}(t)=\omega(A\tau^t(B))$ has an analytic
extension to the complex strip $\{z\mid 0<\mathrm{sign}(\beta)\Im z<|\beta|\}$,
which is bounded and continuous on its closure, and satisfies
$$
F_{A,B}(t+\i\beta)=\omega(\tau^t(B)A)
$$
for all $t\in\RR$. Such states are said to be $(\tau,\beta)$-KMS, and are
$\tau$-invariant.
\end{itemize}

Given such a $C^\ast$-quantum dynamical system, the GNS representation produces
a triple $(\cH,\pi,\Omega)$ where $\cH$ is a Hilbert space, $\pi:\cO\to
\cB(\cH)$ a $\ast$-morphism from $\cO$ to the bounded linear operators on $\cH$,
and $\Omega\in\cH$ a unit vector such that $\omega(A)=\langle\Omega,\pi(A)\Omega\rangle$
for all $A\in\cO$. Moreover, $\Omega$ is cyclic for $\pi(\cO)$, {\sl i.e.,}
$\pi(\cO)\Omega$ is a dense subspace of $\cH$. The weak
closure of the set $\pi(\cO)\subset \cB(\cH)$ coincides with its
bicommutant\footnote{We use the standard notation $\cA'=\{B\in \cB(\cH)\mid
[A,B]=0\text{ for all }A\in\cA\}$ for the commutant of a subset $\cA\subset
\cB(\cH)$.} $\fM=\pi(\cO)''$, and is the {\sl enveloping von Neumann algebra} of
$\cO$ induced by $\omega$. The state $\omega$ clearly extends to a state on
$\fM$ which we denote by the same symbol. A density matrix $\rho$ on $\cH$
defines a state on $\fM$ by the familiar quantum mechanical rule $\fM\ni
A\mapsto\tr(\rho A)$. Such states on $\fM$ are called {\sl normal,} and their
restriction to $\pi(\cO)$ induce states on $\cO$ which are called $\omega$-{\sl
normal.} The {\sl folium} of $\omega$ is the set $\cN$ of all $\omega$-normal
states on $\cO$.

We will always assume that the reference state $\omega$ is {\sl modular\/},
namely that  there exists a $C^\ast$-dynamics $\varsigma_\omega$ on $\cO$ such
that $\omega$ is a $(\varsigma_\omega,-1)$-KMS state. If $\omega$ is
modular, we will say that $(\cO,\tau,\omega)$ is a modular $C^\ast$-quantum
dynamical system. The $C^\ast$-dynamics $\varsigma_\omega$ is the {\sl modular group} of
$\omega$, and is unique when it exists. The extension to $\fM$ of a modular
state is faithful, {\sl i.e.,} the map $\fM\ni A\mapsto A\Omega$ is injective.
It follows from the Tomita--Takesaki  theory, see
{\sl e.g.}~\cite{Derezinski2003a,Jaksic2012} and references therein, that the GNS
Hilbert space $\cH$ comes with a modular structure:
\begin{itemize}
\item A positive operator $\Delta_\omega$, called the {\sl modular operator} of
$\omega$, which implements the modular group on $\pi(\cO)$,
$$
\pi(\varsigma_\omega^\theta(A))=\Delta_\omega^{\i\theta}\pi(A)\Delta_\omega^{-\i\theta},
$$
and thus allows us to extend this group to a $W^\ast$-dynamics on $\fM$,
which we denote by the same symbol.
\item The {\sl modular conjugation} $J$, an anti-unitary involution satisfying
$$
\fM'=J\fM J.
$$
\item The {\sl natural cone,} a self-dual cone $\cH_+\subset\cH$ such that
$\Delta_\omega^{\i\theta}\cH_+=\cH_+$ for all $\theta\in\RR$ and $J\Psi=\Psi$
for all $\Psi\in\cH_+$. Every state $\mu\in\cN$ has a unique vector
representative $\Omega_\mu\in\cH_+$ satisfying
$$
\mu(A)=\langle\Omega_\mu,\pi(A)\Omega_\mu\rangle
$$
for all $A\in\cO$. Moreover, $\Omega_\mu$ is cyclic for $\pi(\cO)$ iff $\mu$
extends to a faithful state on $\fM$, in which case $(\cH,\pi,\Omega_\mu)$ is a
GNS representation induced by $\mu$.
\end{itemize}

A $W^\ast$-dynamics on the enveloping algebra $\fM$ is a group $\RR\ni
t\mapsto\varsigma^t$ of $\ast$-automorphisms of $\fM$ such that the function
$\RR\ni t\mapsto\mu\circ\varsigma^t(A)$ is continuous for all $\mu\in\cN$ and
$A\in\fM$. Given such a $W^\ast$-dynamics $\varsigma$, there exists a unique
self-adjoint operator $\cL$ on the GNS space $\cH$ satisfying
$$
\e^{-\i t\cL}\cH_+=\cH_+,\qquad
\varsigma^t(A)=\e^{\i t\cL}A\e^{-\i t\cL},
$$
for all $t\in\RR$ and $A\in\fM$. In particular, $\e^{-\i t\cL}\Omega_\mu$ is the
vector representative of $\mu\circ\varsigma^t$ in the natural cone. The operator
${\cal L}$ is often called the {\sl standard Liouvillean}. Note that
$\cL_\omega=\log\Delta_\omega$ is the standard Liouvillean of the modular group
$\varsigma_\omega$.

The basic modular structure induced by the modular state $\omega$ is complemented
with:

\begin{itemize}
\item To any pair $(\mu,\nu)$ of elements of $\cN$ is associated another
positive operator $\Delta_{\mu|\nu}$ called {\sl relative modular operator}. We
will consider it only in the case where both $\mu$ and $\nu$ are faithful on
$\fM$. Then $\fM\Omega_\nu$ is a core for $\Delta_{\mu|\nu}^{1/2}$ and one has
$$
J\Delta_{\mu|\nu}^{1/2}A\Omega_\nu=A^\ast\Omega_\mu
$$
for all $A\in\fM$. One checks, see {\sl e.g.}~\cite{Jaksic2012}, that
\beq
\e^{-\i t\cL}\Delta_{\mu|\nu}\e^{\i t\cL}=\Delta_{\mu\circ\tau^t|\nu\circ\tau^t}
\label{eq:DeltaCovar}
\eeq
where $\cL$ is the standard Liouvillean of the $W^\ast$-dynamics $\tau$. We note
also that $\Delta_{\mu|\mu}=\Delta_\mu$, the modular operator of $\mu$, and that
for $\theta\in\RR$ and $A\in\fM$ one has $\Delta_{\mu|\nu}^{\i\theta}A\Delta_{\mu|\nu}^{-\i\theta}
=\Delta_\mu^{\i\theta}A\Delta_\mu^{-\i\theta}=\varsigma_\mu^\theta(A)$.

\item The Araki--Connes cocycle of the pair $(\mu,\nu)$ is the strongly
continuous one parameter family of unitaries  in $\fM$ given by\footnote{Here,
we depart from the traditional convention, denoting by $[D\mu:D\nu]_{\i t}$ what
is usually written $[D\mu:D\nu]_t$.}
\beq
[D\mu:D\nu]_\alpha=\Delta_{\mu|\nu}^\alpha\Delta_\nu^{-\alpha},\qquad(\alpha\in\i\RR).
\label{eq:ConnesDef}
\eeq
This family  satisfies the multiplicative cocycle relation~\cite[Appendix C]{Araki1982}
\beq
[D\mu:D\nu]_{\alpha+\beta}=[D\mu:D\nu]_\alpha\varsigma_\nu^{-\i\alpha}([D\mu:D\nu]_\beta),
\label{eq:gencocycle}
\eeq
and the chain rule
\beq
[D\mu:D\nu]_{\alpha}[D\nu:D\omega]_{\alpha}=[D\mu:D\omega]_{\alpha}
\label{eq:chain}
\eeq
for $\alpha,\beta\in\i\RR$ and $\mu,\nu,\omega\in\cN$.
It also intertwines the modular groups of $\mu$ and $\nu$,
$$
\varsigma_\mu^\theta(A)[D\mu:D\nu]_{\i\theta}=[D\mu:D\nu]_{\i\theta}\varsigma_\nu^\theta(A).
$$

\item Whenever $\mu$ and $\nu$ are faithful on $\fM$, their relative entropy
is defined by
\beq
\Ent(\nu|\mu)=\langle\Omega_\nu,\log\Delta_{\mu|\nu}\Omega_\nu\rangle.
\label{eq:RelEntDef}
\eeq
It satisfies $\Ent(\nu|\mu)\le0$ with equality if and only if $\mu=\nu$.
\end{itemize}

In the algebraic framework, a time-reversal of $(\cO,\tau)$ is an anti-linear
involutive $\ast$-auto\-morphism $\Theta$ of $\cO$ such that
$\Theta\circ\tau^t=\tau^{-t}\circ\Theta$ for all $t\in\RR$. The quantum
dynamical system $(\cO,\tau,\omega)$ is called {\sl time-reversal invariant}
(TRI for short) whenever such a $\Theta$ exists and satisfies
$\omega\circ\Theta(A)=\omega(A^\ast)$ for all $A\in\cO$.

Without further mention, all the $C^\ast$-quantum dynamical systems
$(\cO,\tau,\omega)$ considered in this paper are assumed  to satisfy  the two
basic regularity assumptions of~\cite{Benoist2023a}:

\assuming{Reg1}{RegOne}{%
The family $\{\pi\circ\tau^t\mid t\in\RR\}$  extends to a
$W^\ast$-dynamics on $\fM$ which we again denote by $\tau$. We will denote by $\cL$
its standard Liouvillean.
}

Note that, under this assumption, $\omega_t\in\cN$ for any $t\in\RR$.

\assuming{Reg2}{RegTwo}{%
For all $t\in\RR$ and $\alpha\in\i\RR$,
\[
[D\omega_t:D\omega]_{\alpha}\in\pi(\cO).
\]
}
Whenever the meaning is clear within the context we denote
$\pi^{-1}([D\omega_t:D\omega]_{\alpha})\in \cO$ by
$[D\omega_t:D\omega]_{\alpha}$.

\subsection{Open quantum systems}
\label{sec:repeat-2}

We call \emph{open quantum system} a small system $\sS$, described by a
finite-dimensional Hilbert space $\cH_\sS$, coupled to $M$ thermal reservoirs
$\sR_1,\ldots,\sR_M$. In the algebraic framework, observables of the small
system are elements of the finite-dimensional $C^\ast$-algebra
$\cO_\sS=\cB(\cH_\sS)$. The dynamics $\tau_\sS$ is generated by a self-adjoint
Hamiltonian $H_\sS\in\cO_\sS$,
\beq
\tau_\sS^t(A)=\e^{\i tH_\sS}A\e^{-\i tH_\sS}.
\label{eq:finitesystevol}
\eeq
Each reservoir $\sR_j$ is described by a $C^\ast$-quantum dynamical system
$(\cO_j,\tau_j,\omega_j)$, where $\omega_j$ is a $(\tau_j,\beta_j)$-KMS state
for some $\beta_j>0$. We denote by $\delta_j$ the generator of $\tau_j$.
The joint system $\sS+\sR_1+\cdots+\sR_M$ is described by the $C^\ast$-algebra
$$
\cO=\cO_\sS\otimes\cO_\sR=\cO_\sS\otimes\left(\bigotimes_{j=1}^M\cO_j\right).
$$
The reference state of the joint system is the product state
\beq
\omega=\omega_\sS\otimes\omega_\sR=\omega_\sS\otimes\left(\bigotimes_{j=1}^M\omega_j\right),
\label{eq:omegaOS}
\eeq
where $\omega_\sS$ is an arbitrary $\tau_\sS$-invariant faithful state on
$\cO_\sS$. The state $\omega$ is modular and  its modular group $\varsigma_\omega$
is generated by the $\ast$-derivation\footnote{Whenever the meaning is clear
within the context we write $A$ for $A\otimes\one$ and $\one\otimes A$,
$\delta$ for $\delta\otimes\Id$ and $\Id\otimes\delta$, etc.}
$$
\delta_\omega=\delta_{\omega_\sS}+\delta_{\omega_\sR}
=\i[\log\omega_\sS,\,\cdot\;]-\sum_{j=1}^M\beta_j\delta_j.
$$

The decoupled or free joint dynamics
$$
\tau_\free^t=\tau_\sS^t\otimes\tau_\sR^t
=\tau_\sS^t\otimes\left(\bigotimes_{j=1}^M\tau_j^t\right)
$$
is generated by
$$
\delta_\free=\i[H_\sS,\,\cdot\;]+\sum_{j=1}^M\delta_j,
$$
and commutes with the modular group. Note that since
$\varsigma_{\omega_j}^\theta=\tau_j^{-\beta_j\theta}$, $\tau_j$ satisfies
Assumption~\RegOne, and that its standard Liouvillean is
$\cL_j=-\beta_j^{-1}\log\Delta_{\omega_j}$. One easily infers that $\tau_\free$
also satisfies Assumption~\RegOne, with the standard Liouvillean
$$
\cL_\free=\sum_{j=1}^M\cL_j+H_\sS-JH_\sS J.
$$

The coupling between the small system  and the reservoirs is described by a
self-adjoint element $V\in\cO$ of the form
$$
V=\sum_{j=1}^MV_j,\qquad V_j=V_j^\ast\in\cO_\sS\otimes\cO_j.
$$
The coupled joint system $\sS+\sR_1+\cdots+\sR_M$ is thus described by the $C^\ast$-quantum
dynamical system $(\cO,\tau,\omega)$, where $\tau^t=\e^{t\delta}$ with
$$
\delta=\delta_\free+\i[V,\,\cdot\;].
$$
Invoking time-dependent perturbation theory, the dynamics $\tau$ satisfies
Assumption~\RegOne. Its standard Liouvillean is, see
{\sl e.g.}~\cite{Derezinski2003} and references therein,
\beq
\cL=\cL_\free+V-JVJ.
\label{eq:StdLiouvPert}
\eeq
We note that the self-adjoint operator $\cL_\free+V$ also implements the coupled
dynamics $\tau$ on $\fM$. However, it fails to preserve the natural cone and is
sometimes called semi-standard Liouvillean associated to the local perturbation
$V$.

By~\cite[Lemma~2.4]{Benoist2024b}, Assumption~\RegTwo{} is also satisfied if
$V\in\Dom(\delta_\omega)$ which is equivalent to $V_j\in\Dom(\delta_j)$ for all
$j\in\{1,\ldots,M\}$.

We conclude this section by recalling the definition of Nonequilibrium Steady State (NESS).
This concept was originally introduced in~\cite{Ruelle2000}, and was studied in a number of
follow-up works; an incomplete list of references is~\cite{Ho2000,Pillet2001,
Jaksic2002b,Jaksic2002a,Ruelle2002,Aschbacher2003,Froehlich2003,Matsui2003,Tasaki2003,Ogata2004,
Tasaki2005,Aschbacher2006,Jaksic2006a,Jaksic2006b,Jaksic2006c,Jaksic2006f,Tasaki2006,
Aschbacher2007,Jaksic2007a,Jaksic2007b,Merkli2007,Merkli2007a,Salem2007,
Jaksic2010b,Jaksic2013}. The NESSs of the dynamical system $(\cO,\tau,\omega)$ are
defined as the weak$^\ast$-limit points of the net
\beq
\left\{ \frac{1}{T}\int_0^T \omega_t \d t\,\bigg|\, T>0\right\}
\label{eq:NESS}
\eeq
as $T\to\infty$. The set of NESSs is always non-empty and any NESS is
$\tau$-invariant.

\subsection{Two-time measurement entropy production}
\label{sec-start-te-1}

Consider a finite quantum system with Hamiltonian $H$. Let $\omega$ and $\nu$ be
two  faithful states of this system.
The first state $\omega$ serves as a reference state and
the quantity $S=-\log\omega$ is then the  {\sl entropy observable associated to the reference state $\omega$}.
The second state $\nu$ describes the state
of the system at the instant $t_i$ of a first measurement of $S$. After this
first measurement, whose outcome we denote by $s_i\in\sp(S)$, the system evolves
according to~\eqref{eq:finitesystevol}. At a later time $t_f$ a second
measurement of $S$ is performed with outcome $s_f$. The increment $s=s_f-s_i$ is
interpreted as the entropy produced in system in the time period $t=t_f-t_i$. As
argued in~\cite{Benoist2023a}, the characteristic function of the law
$Q_{\nu,t}$ of the random variable $s$ relates to the modular structure of the
system according to
$$
\int_\RR\e^{-\alpha s}\d Q_{\nu,t}(s)
=\lim_{R\to\infty}\frac1R\int_0^R\nu\circ\varsigma_\omega^\theta
\left([D\omega_{-t}:D\omega]_{\alpha}\right)\d\theta,\qquad(\alpha\in\i\RR).
$$
The interpretation of $s$ as an entropy can be motivated by considering the case of an open
system with finite reservoirs, where the state $\omega$ is given by~\eqref{eq:omegaOS} with
$$
\omega_\sS=\frac{\one}{\tr(\one)},\qquad
\omega_j=\frac{\e^{-\beta_jH_j}}{\tr(\e^{-\beta_jH_j})},
$$
and where $H_j$ denotes the Hamiltonian of the $j^\mathrm{th}$ reservoir.
Then $S=\sum_j\beta_jH_j$ and hence
$$
s=\sum_{j=1}^M\beta_j\Delta E_j,
$$
where $\Delta E_j$ is the measured change in energy of the $j^\mathrm{th}$ reservoir.
Thus, $s$ can be interpreted as the entropy dumped in the reservoirs during the
two-time measurement process.

In the following, we consider the general setting of a $C^\ast$-quantum dynamical
system $(\cO,\tau,\omega)$ with modular reference state $\omega$ and satisfying the
two basic regularity assumptions~\RegOne, \RegTwo.

By~\cite[Theorem 1.3]{Benoist2023a}, for all $\nu\in\cN$, $t\in\RR$ and
$\alpha\in\i\RR$, the limit
\beq
\fF_{\nu,t}^\ttm(\alpha):=\lim_{R\to\infty}\frac1R\int_0^R
\nu\circ\varsigma_\omega^\theta\left([D\omega_{-t}:D\omega]_{\alpha}\right)\d\theta
\label{emm-n}
\eeq
exists, and there is unique Borel probability measure $Q_{\nu,t}^\ttm$ on $\RR$ such that
$$
\fF_{\nu,t}^\ttm(\alpha)=\int_\RR\e^{-\alpha s}\d Q_{\nu,t}^\ttm(s).
$$
Moreover, one also has that
\beq
\fF_{\nu,t}^\ttm(\alpha)=\lim_{R\to\infty}\frac1R\int_0^R
\nu\circ\varsigma_\omega^\theta\left([D\omega_{-t}:D\omega]_{\frac{\bar\alpha}{2}}^\ast
[D\omega_{-t}:D\omega]_{\frac{\alpha}{2}}\right)\d\theta.
\label{toul-late}
\eeq
The measure $Q_{\nu,t}^\ttm$ gives the statistics of an idealized two-time
measurement of the entropy production over a time period of length $t$ in the
system $(\cO,\tau,\omega)$, the latter being in the state $\nu$ at the instant
of the first measurement. The {\sl thermodynamic limit} justification of this
idealization was carried out in~\cite{Benoist2024b}.

As we have already mentioned, in the special case $\nu=\omega$, the family
$(Q_{\omega,t}^\ttm)_{t\in\RR}$ associated to finite quantum systems was
introduced in~\cite{Kurchan2000, Tasaki2000} and was studied in detail
in~\cite{Jaksic2010b}. In the more general setting of algebraic quantum
dynamical systems, it first appeared in~\cite{Tasaki2003}. To the best of our
knowledge, the case of general $\nu\in\cN$ was considered for the first time
in~\cite{ Benoist2023a}, where the following rigidity result was
established~\cite[Theorems 1.5 and 1.6]{ Benoist2023a}.

\begin{theorem}\label{ver-fl-1}
\ben
\item Suppose that the $C^\ast$-quantum dynamical system
$(\cO,\varsigma_\omega,\omega)$ is ergodic. Then for any $\nu\in\cN$
and $t>0$,
\beq
Q_{\nu,t}^\ttm=Q_{\omega,t}^\ttm.
\label{eq:qqequal}
\eeq
\item Suppose that the open quantum system $(\cO,\tau,\omega)$ is such that each
reservoir subsystem $(\cO_j,\tau_j,\omega_j)$ is ergodic. Let $\nu\in\cN$, and
denote by $\nu_\sS$ its restriction to $\cO_\sS$. Then, for all
$\alpha\in\i\RR$,
\beq
\fF_{\nu,t}^\ttm(\alpha)
=\nu_{\sS}\otimes\omega_\sR\left([D{\omega_{-t}}:D{\omega}]_{\alpha}\right)
=\langle\Omega_{\nu_\sS\otimes\omega_\sR},
\Delta_{\omega_{-t}|\omega}^\alpha\Omega_{\nu_\sS\otimes\omega_\sR}\rangle,
\label{eq:ttm-reducedstate}
\eeq
and in particular, $Q_{\nu,t}^\ttm$ is the spectral measure of
$-\log\Delta_{\omega_{-t}|\omega}$ for the vector $\Omega_{\nu_\sS\otimes\omega_\sR}$. Moreover,
if $\nu_\sS$ is faithful, then, for any $t\in\RR$,\footnote{$\sp(\nu_\sS)$ denotes the spectrum of the density matrix associated to the state $\nu_\sS$.}
\beq
\dim(\cH_\sS)\min\sp(\nu_\sS)\leq
\frac{\d Q_{\nu,t}^\ttm}{\d Q_{\omega,t}^\ttm}\leq\dim(\cH_\sS).
\label{est-K-1}
\eeq
\een
\end{theorem}

\preASS{NESS}{NESS}
\noindent{\bf Remark 1.} Theorem~\ref{ver-fl-1}(1) holds in the case of
{\sl directly coupled reservoirs}, {\sl i.e.,} when there is no small system or,
in an equivalent way, when the small system Hilbert space $\cH_\sS$ has
dimension $1$.

\noindent{\bf Remark 2.} In the open quantum system case~(2), the presence of
the small system prevents the validity of~\eqref{eq:qqequal}.
However, Inequalities~\eqref{est-K-1} force the same LDP for all $\nu\in\cN$ (see Remark
after Proposition~\ref{es-thm-1}). Intuitively, \eqref{est-K-1} expresses the fact that,
in the large-time limit, the dynamics of the system is completely dominated by the
reservoirs.

\noindent{\bf Remark 3.} Recall that, by~\RegOne, $\omega_T\in\cN$ for all $T\in\RR$, so that
the previous theorem applies to $\nu=\omega_T$. $Q^\ttm_{\omega_T,t}$ can also be
interpreted as corresponding to a measurement protocol where the system, initially in state
$\omega$, is measured first at time $T$ and then at time $T+t$. In this case, the equality
$Q^\ttm_{\omega_T,t}=Q^\ttm_{\omega,t}$ and the Inequalities~\eqref{est-K-1} can be interpreted as the memory erasing effect
due to the decoherence induced by the first measurement on the reservoirs.

\medskip
Denote by $\cP(\RR)$ the set of all Borel probability measures on $\RR$ equipped
with the weak topology. By~\cite[Lemma~2.1]{Takenouchi1955}
and~\cite[Theorem~1.1]{Fell1960}, see also~\cite[Theorem~2.2.13]{Haag1996}, the
folium $\cN$ is dense in $\cS_\cO$. Hence, under the assumptions of
Theorem~\ref{ver-fl-1}, Part~(1) or~(2), the map
\[
\cN\ni\nu\mapsto Q_{\nu,t}^\ttm\in\cP(\RR)
\]
uniquely extends to a continuous map
\[
\cS_\cO\ni\nu\mapsto Q_{\nu,t}^\ttm\in\cP(\RR).
\]
If Part~(1) holds then obviously $Q_{\nu,t}^\ttm=Q_{\omega,t}^\ttm$ for all $\nu\in\cS_\cO$.
In the case of open quantum systems $Q_{\nu,t}^\ttm$ is again the spectral measure
of $-\log\Delta_{\omega_{-t}|\omega}$ for the vector $\Omega_{\nu_\sS\otimes\omega_\sR}$,
and the estimates~\eqref{est-K-1} hold if $\nu_\sS$ is faithful. This defines the 2TMEP
of $(\cO,\tau,\omega)$ with respect to any initial state $\nu\in\cS_\cO$, and
applies, in particular, to a NESS $\omega_+$ as defined in~\eqref{eq:NESS}.
Under additional hypothesis, $Q_{\omega_+,t}^\ttm$ can be obtained as the weak limit
of $Q_{\omega_T,t}^\ttm$ as $T\to\infty$, see Assumption~\preNESS{} below.

The above  results  resolve the obstacle~(a) mentioned in the introduction.
However, this achievement has been obtained in the context of idealized
measurements. Indeed, the ergodicity assumptions of Theorem~\ref{ver-fl-1} require
the system under consideration, or the reservoirs in the case of an
open system, to be infinitely extended. This means in particular
that the energies supposed to be measured are infinite\footnote{On a more
technical level, the quantity $S$ is no more an observable.}. In the spirit of
statistical mechanics, such idealized measurements can be understood as approximating
those made on a large but finite system. We refer the reader to~\cite{Benoist2023a}
for further discussion  and to~\cite{Benoist2024b} for a mathematical justification
of this {\sl thermodynamic limit.}

\subsection{Entropic ancilla state tomography}
\label{sec-start-te-2X}

To provide another interpretation of the probability measure
$Q_{\omega,t}^\ttm$, and overcome obstacle~(b), we shall couple the system
$(\cO,\tau,\omega)$ to an auxiliary finite quantum system, called the {\sl
ancilla}, and replace the two idealized measurements of the entropic observable
$S$ with a specific sequence of projective measurements performed on the
ancilla, a procedure often called tomography of the ancilla state. We shall see
that, under appropriate choice of the coupling to a single qubit, one can relate
the state of this ancillary qubit to the functional
\beq
\fF_{ \nu,t}^\anc(\alpha):=\nu\left([D\omega_{-t}:D\omega]^\ast_{\frac{\bar \alpha}{2}}
[D\omega_{-t}:D\omega]_{\frac{\alpha}{2}}\right),
\label{sn-mon-1}
\eeq
where $\nu$ is an arbitrary initial state of the system.\footnote{Note that, for
$\nu\neq\omega$, $\fF_{ \nu,t}^\anc$ is not necessarily the Fourier transform of
a Borel probability measure.} It follows from~\eqref{toul-late} that
$\fF_{\omega,t}^\ttm=\fF_{\omega,t}^\anc$, but for an arbitrary initial state
$\nu$ the functionals $\fF_{\nu,t}^\ttm$ and $\fF_{\nu,t}^\anc$ will be
generally distinct. However, when $\nu$ is chosen to be the NESS, the asymptotic
relation between $\fF_{\nu,t}^\ttm$ and $\fF_{\nu,t}^\anc$, in the limit
$t\to\infty$, is part of the PREF, see Definition~\ref{def-sqprf-1}. From the
physics perspective the main point is that the ancilla's state, and hence the
functional $\fF_{\nu,t}^\anc$, are experimentally accessible through state
tomography. We refer the interested reader
to~\cite{Dorner2013,Campisi_2013,Johnson2016,Roncaglia2014,Chiara2015,
Goold2014,Mazzola2013} for related theoretical studies in the physics literature
and to~\cite{An2014,Batalhao2014,Batalhao2015,Peterson2019} for experimental
implementations.

To elucidate the definition of $\fF_{\nu,t}^\anc$, let us consider a
finite quantum dynamical system $(\cO,\tau,\omega)$ where:
\begin{itemize}
\item $\cO=\cB(\cK)$ for a finite-dimensional Hilbert space $\cK$.
\item $\tau$ is the $C^\ast$-dynamics generated by the self-adjoint Hamiltonian $H\in\cO$.
\item $\omega$ is a faithful density matrix on $\cO$.
\end{itemize}
In this case it follows that
$$
[D\omega_{-t}:D\omega]_{\frac{\alpha}{2}}=\omega_{-t}^{\alpha/2}\omega^{-\alpha/2},
$$
and  Definition~\eqref{sn-mon-1} becomes
$$
\fF_{ \nu,t}^\anc(\alpha)=\tr(\nu\omega^{-\alpha/2}\omega_{-t}^\alpha\omega^{-\alpha/2}).
$$

The ancilla's Hilbert space is $\CC^2$ and its initial state is a density matrix
\[
\rho =\begin{bmatrix}
\rho_{++}&\rho_{+-}\\[2pt]
\rho_{-+}&\rho_{--}
\end{bmatrix},
\]
where $\rho_{+-}\not=0$.
The Hilbert space of the coupled system is $\widehat\cK=\cK\otimes\CC^2$
and its initial state is $\widehat\nu=\nu\otimes\rho$, where $\nu$ is
a density matrix on $\cK$. The coupling between the system and the ancilla
is given by the Hamiltonian
$$
\widehat H_\alpha=\e^{\frac\alpha2\log\omega\otimes\sigma_z}
\left(H\otimes\one\right)\e^{-\frac\alpha2\log\omega\otimes\sigma_z},
$$
parametrized by $\alpha\in\i\RR$. A simple calculation gives that the ancilla's
state at time $t$ is given by
\beq
\rho_t=\tr_\cK(\e^{-\i t\widehat H_\alpha}\widehat \nu\e^{\i t\widehat H_\alpha})
=\begin{bmatrix}
\rho_{++}&\fF_{ \nu,t}^\anc(\alpha)\rho_{+-}\\[2pt]
\overline{\fF_{ \nu,t}^\anc(\alpha)}\rho_{-+}&\rho_{--}
\end{bmatrix}.
\label{eq:wellfare}
\eeq
Note that in the special case $H=H_\free+V$ with $[\omega,H_\free]=0$%
\footnote{This will be the case in finite open quantum systems.},
one has $\widehat H_\alpha=H\otimes\one+\widehat W_\alpha$ where
\beq
\widehat W_\alpha=\tfrac12W_\alpha\otimes(\one+\sigma_z)+\tfrac12W_{-\alpha}\otimes(\one-\sigma_z),
\label{tuluz-1}
\eeq
with
\beq
W_\alpha=\varsigma_\omega^{-\i\alpha/2}(V)-V.
\label{tuluz-2}
\eeq

To describe the general case, we denote by $\Mat_2(\CC)$ the algebra of complex
$2\times2$ matrices and consider an open quantum system $(\cO,\tau,\omega)$ with
coupling $V$. The algebra of observables of this system coupled to a qubit is
$\widehat\cO:=\cO\otimes{\Mat_2(\CC)}$ which we will often identify with the
algebra $\Mat_2(\cO)$ of $2\times2$ matrices with entries in $\cO$,
see~\cite[Section 2.7.2]{Bratteli1987}. The decoupled dynamics on $\widehat\cO$
is given by $\widehat\tau^t=\tau^t\otimes\Id$ while $W_\alpha$ and $\widehat
W_\alpha$ are defined by~\eqref{tuluz-1} and~\eqref{tuluz-2} (note that  they
are self-adjoint). Let
$$
\widehat\tau_\alpha^t=\e^{t\hat\delta_\alpha},\qquad
\widehat\delta_\alpha=\delta\otimes\Id+\i[\widehat{W}_\alpha,\,\cdot\;],
$$
be the perturbation of $\widehat\tau$ by $\widehat W_\alpha$ and let $\rho$ be
as above. For $\nu\in\cS_\cO$ we set $\widehat\nu=\nu\otimes\rho$,
$\widehat\nu_t=\widehat\nu\circ\widehat\tau^t$, and
$\rho_t=\widehat\nu_t|_{\Mat_2(\CC)}$. We then have:

\bep\label{ancilla-start-1}
Suppose that $V\in\Dom(\delta_\omega)$. Then, for all $\alpha \in \i\RR$ and $\nu\in\cS_\cO$, the
ancilla's state at time $t\in\RR$ is given by the right-hand side of~\eqref{eq:wellfare}.
\eep

The proof is given in Section~\ref{sec-proof-ancilla-1}.

Beyond open quantum systems, the definition~\eqref{sn-mon-1} remains useful in
the study of the general mathematical structure of non-equilibrium quantum
statistical mechanics, and we will make use of it in that context.

The identity $\fF_{\omega,t}^\ttm=\fF_{\omega,t}^\anc$ is of considerable
theoretical and practical importance. The two-time measurement entropy
production protocol is always introduced in the context of finite quantum
systems (or, slightly more generally, confined quantum system with possibly
infinite discrete energy spectra). This identity is also of experimental
relevance. Indeed, the ancilla technique has been used to access the two time
measurement distribution of work~\cite{Cerisola2014,Cerisola2017,DeChiara2018}.

Thanks to its connection with modular theory, the two-time measurement
statistics has a thermodynamic limit~\cite{Benoist2024b} which, by the identity
$\fF_{\omega,t}^\ttm=\fF_{\omega,t}^\anc$, is experimentally accessible through
ancilla state tomography. This relation makes it possible to interpret the
results of entropic ancilla state tomography in terms of energy transfers in
open quantum systems with large reservoirs.

\subsection{The quantum principles of regular entropic fluctuations}
\label{sec-quantum-pref}

We recall the basic properties of $Q_{\omega, t}^\ttm$;
see~\cite[Theorem~7]{Tasaki2003} and~\cite[Theorem~1.4]{Benoist2023a}.

\bep\label{gen-ttm-quant}
\ben
\item $\int_\RR s\,\d Q_{\omega,t}^\ttm(s)=-\Ent(\omega_t|\omega)$.
In particular,
$$
\int_\RR s\,\d Q_{\omega,t}^\ttm(s)\geq 0,
$$
with equality iff $\omega=\omega_t$.
\item The map $\i\RR\ni\alpha\mapsto\fF_{\omega,t}^\ttm(\alpha)$ has an analytic
extension to the vertical strip $0<\Re\alpha<1$ which is bounded and continuous
on its closure.
\een

In the remaining statements we assume that $(\cO,\tau,\omega)$ is time-reversal
invariant.

\ben\setcounter{enumi}{2}
\item For any $\alpha$ satisfying $0\leq \Re\,\alpha\leq 1$,
$$
\fF_{\omega, t}^\ttm(\alpha)={\overline{ \fF_{\omega, t}^\ttm(1-\bar \alpha)}}.
$$
\item Let $\mathfrak{r}:\RR\rightarrow \RR$ be the reflection at $0$,
$\mathfrak{r}(s)=-s$, and $\bar
Q_{\omega,t}^\ttm=Q_{\omega,t}^\ttm\circ\mathfrak{r}$. Then the measures
$Q_{\omega,t}^\ttm$ and ${\bar Q}_{\omega,t}^\ttm$ are mutually absolutely
continuous and
\beq
\frac{\d\bar Q_{\omega,t}^\ttm}{\d Q_{\omega, t}^\ttm}(s)=\e^{-s}.
\label{str-2-quant-1}
\eeq
\een
\eep

We consider the family $(P_{\omega,t}^\ttm)_{t>0}\subset\cP(\RR)$ defined by
\[
P_{\omega,t}^\ttm(B)=Q_{\omega,t}^\ttm(tB),
\]
for all Borel sets $B\subset\RR$ and $t>0$. It describes the statistics of
the two-time measurement entropy production {\sl per unit time} of
$(\cO,\tau,\omega)$ with respect to $\omega$ over the time interval $[0,t]$.
The relation~\eqref{str-2-quant-1} has an important consequence for
the Large Deviation Principle (LDP) satisfied by $(P_{\omega,t}^\ttm)_{t>0}$.
Before describing it, we give a short general overview
of LDP that is suited for our purposes.

\begin{definition}
The family $(P_t)_{t>0}\subset\cP(\RR)$ satisfies a \textbf{full LDP} if there
exists a lower-semicontinuous function $\II:\RR\to[0,\infty]$, called the
\textbf{rate function}, such that for any Borel set $B\subset \RR$,
\beq
-\inf_{s\in\interior(B)}\II(s)\leq \liminf_{t\to\infty}\tfrac1t\log P_t(B)
\leq \limsup_{t\to\infty}\tfrac1t\log P_t(B)\leq -\inf_{s\in\closure(B)}\II(s),
\label{ldp-rate-11}
\eeq
where $\interior(B)/\closure(B)$ denotes the interior/closure of $B$.

If~\eqref{ldp-rate-11} holds only for Borel sets $B\subseteq\, ]a,b[$,
where $]a, b[\not=\RR$, we then say that a \textbf{local LDP}
holds for $(P_t)_{t>0}$ on the interval $]a,b[$.
\end{definition}

\noindent
{\bf Remark.} The lower-semicontinuity assumption ensures that the rate function
$\II$ is unique whenever it exists.

\medskip
The celebrated  Gärtner--Ellis theorem gives an important criterion that ensures
the validity of the LDP.

\begin{theorem}\label{thm:GEapplied}
Let $I=]\vartheta_-,\vartheta_+[\subset\RR$ be an open interval containing $0$, and
suppose that the limit
\[
F(\alpha)=\lim_{t\to\infty}\frac{1}{t}\log \int_\RR e^{-\alpha s}\d P_t(s)
\]
exists,  is finite for $\alpha\in I$, and that the function  $F:I\to\RR$ is
differentiable. If $I=\RR$, then the full LDP holds with the rate function
\beq
\II(s)=\sup_{-\alpha\in I}(s\alpha-F(-\alpha)).
\label{ldp-rate-21}
\eeq
Otherwise, the local LDP with rate function~\eqref{ldp-rate-21} holds for any
Borel $B\subset]a, b[$ where
\beq
a=\lim_{\alpha\downarrow\vartheta_-}F^\prime(\alpha),\qquad
b=\lim_{\alpha\uparrow\vartheta_+}F^\prime(\alpha).
\label{ldp-limits-1}
\eeq
\end{theorem}

\noindent
{\bf Remark.} By Hölder's inequality, $F$ is convex. Thus,
$F^\prime$ is increasing on $I$ and the limits~\eqref{ldp-limits-1} always exist.
It may happen that $a=-\infty$ and $b=\infty$ (even for bounded $I$), in which case the
full LDP again holds. Otherwise, if either $a$ or $b$ is finite, the Gärtner--Ellis
theorem yields only a local LDP. For more information about LDP we refer the reader
to~\cite{Dembo2000,Ellis2006}.

\medskip
Returning to the family $(P_{\omega,t}^\ttm)_{t>0}$, a consequence of the relation
\eqref{str-2-quant-1} is:

\bep\label{es-thm-1}
Suppose that the system $(\cO,\tau,\omega)$ is time-reversal invariant and that the full LDP holds
for the family $(P_{\omega, t}^\ttm)_{t>0}$ with rate function $\II$. Then for all $s\in \RR$,
\beq
\II(-s)= \II(s) + s.
\label{quantum-es-1}
\eeq
If the local LDP holds on $]-a, a[$ for some $a>0$, then~\eqref{quantum-es-1} holds for $s\in]-a,a[$.
\eep

\noindent{\bf Remark.} Under the hypothesis of Theorem~\ref{ver-fl-1}, one has obviously
$$
\lim_{t\to\infty}\frac1t\log\fF^\ttm_{\omega,t}(\alpha)
=\lim_{t\to\infty}\frac1t\log\fF^\ttm_{\nu,t}(\alpha)
$$
for all $\nu\in\cN$. Therefore, whenever Theorem~\ref{thm:GEapplied} implies the
LDP for $(P^\ttm_{\omega,t})_{t>0}$, the {\sl same} LDP must hold for $(P^\ttm_{\nu,t})_{t>0}$, with
the {\sl same} rate function. Notice that this holds in particular for $\nu=\omega_T$.

\proof We follow~\cite{Cuneo2017} and prove the result in the full LDP case. The
local LDP case is identical. We abbreviate $Q_{\omega,t}^\ttm$ and $P_{\omega,t}^\ttm$
with $Q_t$ and $P_t$. Relation~\eqref{str-2-quant-1} gives that for any
Borel set $B\subset\RR$ we have
\[
Q_t(B)\leq\e^{\sup B}Q_t(-B).
\]
Replacing $B$ with $tB$, the LDP gives
\[
-\inf_{u\in\interior(B)}\II(u)\leq\liminf_{t\to\infty}\tfrac1t\log P_t(B)
\leq\limsup_{t\to\infty}\tfrac1t\log\left(\e^{t\sup B}P_t(-B)\right)\leq\sup B-\inf_{u\in\closure(B)}\II(-u).
\]
Taking $B=]s-\epsilon, s+\epsilon[$ we derive
\beq
\inf_{|u+s|<2\epsilon}\II(u)
\leq \inf_{|u+s|\leq \epsilon}I(u)
\leq s+\epsilon + \inf_{|u-s|<\epsilon}\II(u).
\label{sat-ajde-q-1}
\eeq
Since the function $\II$ is lower semicontinuous,
\[
\II(s)=\lim_{\epsilon\downarrow 0}\inf_{|u-s|<\epsilon} \II(u),
\]
and~\eqref{sat-ajde-q-1} gives that $\II(-s)\leq s+\II(s)$ for any $s\in\RR$.
Replacing $s$ with $-s$ and combining the two inequalities we
derive~\eqref{quantum-es-1}. \hfill\qed

Due to the obvious parallel with the foundational works of~\cite{Evans1994} in
classical statistical mechanics\footnote{See~\cite{Jaksic2011} and
Section~\ref{sec-classical-1} below.}, if the full/local LDP holds for
$(P_{\omega, t}^\ttm)_{t>0}$, we will say that the full/local quantum
Evans--Searles fluctuation theorem holds for $(\cO,\tau,\omega)$. The
relations~\eqref{str-2-quant-1} and~\eqref{quantum-es-1} are sometimes called
quantum Evans--Searles fluctuation relation. We emphasize
that~\eqref{str-2-quant-1} is an immediate consequence of time-reversal
invariance, and that~\eqref{quantum-es-1} follows from~\eqref{str-2-quant-1} and
the LDP.

The equally celebrated Gallavotti--Cohen fluctuation
theorem~\cite{Gallavotti1995b,Gallavotti1995c} refers to the
statistics of entropy production with respect to the non-equilibrium steady state
reached in the long-time limit. In spite of a formal similarity, it is
conceptually and technically a very different statement than the Evans--Searles
theorem that refers to the statistics of the entropy production with respect to
the reference (initial) state of the system. The two theorems are related by an
exchange of limits. The validity of this exchange of limits is a deep dynamical
problem that has been lifted to the principle of regular entropic fluctuations
in~\cite{Jaksic2011}. We will review these points in
Section~\ref{sec-classical-1} and~\ref{sec-cl-q-comp}; for an in depth
discussion see~\cite{Jaksic2011}.

Returning to quantum statistical mechanics, we make the following assumption
concerning the NESS defined in~\eqref{eq:NESS}:

\assuming{NESS}{NESS}{%
For all $A\in\cO$, the limit
\beq
\lim_{t\to\infty}\omega\circ\tau^t(A)=\omega_+(A)
\label{paris-so}
\eeq
exists, so that $\omega_+$ is the unique NESS of the
system $(\cO,\tau,\omega)$. Moreover, for all $t>0$ the weak limit
\beq
Q_{\omega_+,t}^\ttm:=\lim_{T\to\infty}Q_{\omega_T,t}^\ttm
\label{so-so-ht-1}
\eeq
exists.\footnote{Recall that~\eqref{paris-so} $\Rightarrow$~\eqref{so-so-ht-1}
under the assumptions of Theorem~\ref{ver-fl-1}.}
}

The family $(P_{\omega_+, t}^\ttm)_{t>0}$ is defined by
$P_{\omega_+,t}^\ttm(B)=Q_{\omega_+, t}^\ttm(tB)$. In parallel with the
classical theory of entropic fluctuations~\cite{Jaksic2011}, see also
Section~\ref{sec-classical-1}, the following definitions are natural.

\begin{definition} Suppose that \NESS{} holds. We say that the \textbf{full weak
quantum Gallavotti--Cohen theorem} holds for $(\cO,\tau,\omega)$ if
$(P_{\omega_+,t}^\ttm)_{t>0}$ satisfies the full LDP. The \textbf{local weak
quantum Gallavotti--Cohen theorem} holds if the LDP holds on some finite
interval $]-a, a[$, $a>0$.
\end{definition}

\begin{definition} Suppose that \NESS{} holds. We say that the \textbf{full weak
quantum PREF} holds for $(\cO,\tau,\omega)$ if the families
$(P_{\omega,t}^\ttm)_{t>0}$ and $(P_{\omega_+,t}^\ttm)_{t>0}$ both satisfy the
full LDP with the same rate function. The \textbf{local weak quantum PREF} holds
if they satisfy local LDP on the same finite interval $]-a,a[$, $a>0$, with the
same rate function.
\end{definition}

In the context of 2TTM, the PREF essentially trivializes.
Indeed, in the directly coupled case, Theorem~\ref{ver-fl-1}(1) implies immediately
$Q_{\omega_+,t}=Q_{\omega,t}$, while in the open quantum system case (2), it suffices
to notice that the function $F(\alpha)$ in Theorem~\ref{thm:GEapplied} is the
same whenever computed with respect to $\omega$ or $\omega_+$ (see remark after
Proposition~\ref{es-thm-1}). Therefore, the LDP for $(P^\ttm_{\omega,t})_{t>0}$
implies the {\sl same} LDP for $(P^\ttm_{\omega_+,t})_{t>0}$.

Again this trivialization can be ultimately understood as a consequence of the
first measurement's dominating effect in the thermodynamic limit. This rigidity
has no classical analog. In the classical case, the equality of the rate
functions is not only non-trivial, but it can be regarded as a deep dynamical
property of the system.

However, in a surprising turn, the parallel with the classical PREF is
restored when following another route to the quantum extension of entropic
functionals, given by EAST, which we now describe. In order to open this route,
we need to introduce an additional regularity assumption which will allow to
extend the entropic functionals to a complex domain.

To $\vartheta >0$ we associate the vertical strip
\[
\fS(\vartheta)=\left\{ z\in \CC\mid|\Re\, z| <\vartheta\right\},
\]
and the assumption

\assuming{AnC$\boldsymbol{(\vartheta)}$}{AnC}{%
For any $t\in\RR$ the function
\[
\i \RR\ni \alpha \mapsto [D\omega_{t}: D\omega]_\alpha \in \cO
\]
has an analytic extension to the strip $\fS(\vartheta)$.
}

The cocycle relation~\eqref{eq:gencocycle} gives
$$
[D\omega_t: D\omega]_{\alpha_1+\alpha_2}
=[D\omega_t: D\omega]_{\alpha_1}\varsigma_\omega^{-\i\alpha_1}([D\omega_t: D\omega]_{\alpha_2}),
$$
which holds for $\alpha_1,\alpha_2\in\i\RR$. Assuming \AnC, this last relation
extends by analytic continuation to $\alpha_1\in\i\RR$ and
$\alpha_2\in\fS(\vartheta)$. This gives that if \AnC{} holds, then for all
$\alpha$ in the sub-strip $\fS(\vartheta^\prime)$,
$0<\vartheta^\prime<\vartheta$,
\beq
\|[D\omega_s:D\omega]_\alpha\|
\leq\sup_{|\gamma|<\vartheta^\prime}\|[D\omega_s:D\omega]_\gamma\|.
\label{mar-hun}
\eeq
The next proposition is an immediate consequence of this bound and Vitali's convergence
theorem~\cite[Theorem~5.21]{Titchmarsh1939} applied to Relation~\eqref{emm-n}.

\bep\label{proja-1}
Suppose that Assumptions~\NESS{} and~\AnC{} hold. Then for any $T>0$ and $t\in\RR$ the map
\[
\i\RR\ni\alpha\mapsto\fF_{\omega_T,t}^\ttm(\alpha)
\]
has an analytic continuation to the strip $\fS(\vartheta)$ such that, for any
$\alpha$ in this strip, the limit
\[
\fF_{\omega_+, t}^\ttm(\alpha):=\lim_{T\to\infty}\fF_{\omega_T, t}^\ttm(\alpha)
\]
exists and is finite. Moreover, the function $\alpha\mapsto\fF_{\omega_+,t}^\ttm$
is analytic on $\fS(\vartheta)$, and for any $\alpha$ in this strip,
\beq
\fF_{\omega_+,t}^\ttm(\alpha)=\int_\RR\e^{-\alpha s}\d Q_{\omega_+,t}^\ttm(s).
\label{eq:ttmnesslaplace}
\eeq
\eep

\noindent
{\bf Remark.} By~\eqref{eq:ttmnesslaplace}, $\fF_{\omega_+,t}^\ttm$ obviously extends
analytically to the half-plane $\Re(\alpha)>0$, but we will not make use of that fact.

\medskip

It follows directly from \NESS{} and \AnC{} that for all $t>0$ the function
$\i\RR\ni\alpha\mapsto\fF_{\omega_+,t}^\anc(\alpha)$ has an analytic continuation
to the strip $\fS(2\vartheta)$, and that for $\alpha$ in this strip,
\[
\fF_{\omega_+,t}^\anc(\alpha)=\omega_+\left([D\omega_{-t}:D\omega]^\ast_{\frac{\bar\alpha}{2}}
[D\omega_{-t}:D\omega])_{\frac{\alpha}{2}}\right).
\]

\begin{definition} We say that $(\cO,\tau,\omega)$ satisfies the \textbf{strong quantum PREF}
on the interval $]\vartheta_-, \vartheta_+[$ containing $0$ if Assumptions \NESS{}
and \AnC{} hold, with $\vartheta>\max\{|\vartheta_-|,\vartheta_+\}$, and the limits
 \begin{align*}
\bF^\ttm_{\omega}(\alpha)&:=\lim_{t\to\infty}\tfrac1t\log\fF_{\omega, t}^\ttm(\alpha),\\[1mm]
\bF^\ttm_{\omega_+}(\alpha)&:=\lim_{t\to\infty}\tfrac1t\log\fF_{\omega_+, t}^\ttm(\alpha),\\[1mm]
\bF^\anc_{\omega_+}(\alpha)&:=\lim_{t\to\infty}\tfrac1t\log\fF^\anc_{\omega_+, t}(\alpha),
\end{align*}
exist for all $\alpha\in]\vartheta_-,\vartheta_+[$, and define differentiable functions
on this interval satisfying
\beq
\bF^\ttm_{\omega}=\bF^\ttm_{\omega_+}=\bF^\anc_{\omega_+}.
\label{go-w-1}
\eeq
\label{def-sqprf-1}
\end{definition}

\noindent
{\bf Remark.} Setting
\[
a=\lim_{\alpha\downarrow\vartheta_-}\partial_\alpha\bF_{\omega}^\ttm(\alpha), \qquad
b=\lim_{\alpha\uparrow\vartheta_+}\partial_\alpha\bF_{\omega}^\ttm(\alpha),
\]
and invoking the Gärtner--Ellis theorem, the strong quantum PREF implies that
the families $(P_{\omega, t}^\ttm)_{t>0}$ and $(P_{\omega_+, t}^\ttm)_{t>0}$
both satisfy a LDP on the interval $]a,b[$, with the same rate function.
If either $]\vartheta_-,\vartheta_+[\,=\RR$ or $]a,b[=\RR$, we say that the
full strong quantum PREF holds, otherwise that the local strong quantum PREF holds.
Obviously, the local/full strong quantum PREF implies the local/full weak quantum PREF.

\medskip
If the assumptions of Theorem~\ref{ver-fl-1} are satisfied, either~(1) or~(2),
then~\eqref{so-so-ht-1} and the first equality in~\eqref{go-w-1} auto\-ma\-ti\-cally
hold. We shall see that under the following regularity assumption, which will
also play an important role in our discussion  of quantum transfer operators,
the second relation in~\eqref{go-w-1} can be reduced to an exchange of limits.

\assuming{AnV$\boldsymbol{(\vartheta)}$}{AnV}{%
$(\cO, \tau, \omega)$ describes an open quantum system
where the connecting perturbation $V$ is such that the map
\[
\i\RR\ni\theta\mapsto\varsigma_\omega^{-\i\theta}(V)\in\cO
\]
has an analytic extension to the strip $\fS(\vartheta)$.
}

\bep\label{prop-qto-1}
Suppose that \AnV{} holds. Then \AnC{} holds, and for any $t\in\RR$ and
$\alpha\in\fS(\vartheta)$
\beq
\|[D\omega_t:D\omega]_\alpha\|\leq\e^{|t|(\|\varsigma_\omega^{-\i\Re\,\alpha}(V)\| +\|V\|)}.
\label{est-sa}
\eeq
\eep

\bep\label{te-pa-1}
Suppose that \AnV{} holds for some $\vartheta>\frac12$ and set
\begin{align*}
C_T&:=\e^{2|T|(\|\varsigma_\omega^{-\i/2}(V)\|+\|V\|)},\\[1mm]
D_T&:=\e^{-2|T|(\|\varsigma_\omega^{\i/2}(V)\|+\|V\|)}.
\end{align*}
Then for any $\alpha\in]-\vartheta,\vartheta[$,
\beq
 D_T\fF_{\omega, t}^\ttm(\alpha)\leq \fF_{\omega_T, t}^\anc(\alpha)\leq C_T\fF_{\omega, t}^\ttm(\alpha).
\label{proja-2}
\eeq
\eep

\noindent
{\bf Remark 1.} Assuming the existence of $\bF_{\omega}^\ttm$, the estimate~\eqref{proja-2}
gives that for $\alpha\in]-\vartheta,\vartheta[$,
\[
\bF_{\omega}^\ttm(\alpha)=\lim_{T\to\infty}\lim_{t\to\infty}\tfrac1t\log\fF_{\omega_T,t}^\anc(\alpha).
\]
It follows that the second relation in~\eqref{go-w-1} holds iff the exchange of limits
\beq
\lim_{T\to\infty}\lim_{t\to\infty}\tfrac1t\log\fF_{\omega_T,t}^\anc(\alpha)
=\lim_{t\to\infty}\lim_{T\to\infty}\tfrac1t\log\fF_{\omega_T, t}^\anc(\alpha)
\label{qu-ex-lim-1}
\eeq
is valid for $\alpha\in]-\vartheta, \vartheta[$.

\noindent
{\bf Remark 2.} Under the assumptions of Proposition~\ref{te-pa-1}, for $\alpha\in]-\vartheta,\vartheta[$
we also have the estimates
\beq
D_T\fF_{\omega, t }^\ttm(\alpha)\leq \fF_{\omega_T, t}^\ttm(\alpha)\leq C_T\fF_{\omega, t}^\ttm(\alpha).
\label{t-na-1}
\eeq
Theorem~\ref{ver-fl-1} of course provides much stronger estimates with
$T$-independent constants, but they hold only under  ergodicity assumptions. The
estimates~\eqref{t-na-1} only require the regularity assumption \AnV. In
particular, they hold for finite quantum systems to which Theorem~\ref{ver-fl-1}
cannot be applied.

\medskip
Proposition~\ref{prop-qto-1} is proven in Section~\ref{sec-proof-gto-1} while
Proposition~\ref{te-pa-1} and Estimate~\eqref{t-na-1} are proven in
Section~\ref{sec-proof-te-pa}.

The validity of an exchange of limits plays an equally crucial role in the
theory of classical PREF, see~\eqref{cl-ex-lim-1}. For that and other comparison
reasons, we review briefly the classical theory before returning to the quantum
case.

\subsection{Entropy production and entropic fluctuations in classical dynamical systems}
\label{sec-classical-1}

This section follows~\cite{Jaksic2011}. Since this material was not
discussed in~\cite{ Benoist2023a} we provide some details, referring the reader
to~\cite{Jaksic2011} for a complete exposition and proofs.

We start with a pair $(\cX,\phi)$, where $\cX$ is a compact metric space and
$\phi=\{\phi^t\mid t\in\RR\}$ is a group of homeomorphisms of $\cX$ such that
the map
\[
\RR\times\cX\ni(t,x)\mapsto\phi^t(x)\in\cX
\]
is continuous. We denote by $C(\cX)$ the vector space of all continuous
complex-valued functions on $\cX$ and equip it with the sup norm
$\|f\|_{\infty}:=\sup_{x\in\cX}|f(x)|$. Observables are functions $f\in C(\cX)$,
and they evolve in time as $f\mapsto f_t=f\circ\phi^t$. States are Borel
probability measures on $\cX$, and we write $\nu(f)=\int_\cX f\d\nu$. They
evolve in time as $\nu \mapsto\nu_t=\nu\circ\phi^{-t}$. A state $\nu$ is
$\phi$-invariant if $\nu_t=\nu$ for all $t$. The relative entropy of two states
$\nu$ and $\mu$ is defined by
\[
\Ent(\nu|\mu)=
\begin{cases}
\int_\cX\log\left(\frac{\d\mu}{\d\nu}\right)\d\nu&\text{if }\mu\ll\nu;\\[4pt]
-\infty&\text{otherwise}.
\end{cases}
\]
Its basic property is $\Ent(\nu|\mu)\leq0$ with equality iff $\nu=\mu$.

A time-reversal of $(\cX,\phi)$ is an involutive homeomorphism
$\iota:\cX\to\cX$ such that
\[
\iota\circ\phi^t=\phi^{-t}\circ\iota
\]
for all $t\in \RR$. Given such a time-reversal map $\iota$, a state $\nu$
is called time-reversal invariant (TRI) if $\nu\circ\iota=\nu$.

Our starting point is a classical dynamical system $(\cX,\phi,\omega)$ where, to avoid
triviality, the reference (initial) state $\omega$ is supposed not to be $\phi$-invariant.
The system is TRI if $\omega$ is TRI for some time-reversal of $(\cX,\phi)$.

We set the regularity assumptions. The first one is:
\assuming{Cl1}{ClOne}{%
For all $t\in \RR$ the measures $\omega$ and $\omega_t$ are mutually absolutely continuous.
}

This allows us to set
\[
\Delta_{\omega_t|\omega}:=\frac{\d\omega_t}{\d\omega},
\qquad \ell_{\omega_t|\omega}:=\log \Delta_{\omega_t|\omega},
\qquad c^t:=\ell_{\omega_t|\omega}\circ \phi^t= \ell_{\omega|\omega_{-t}},
\]
for $t\in\RR$. The following property is immediate and will play a central role.

\bep\label{chain-cl} The family $(c^t)_{t\in\RR}$ is an additive $\phi$-cocycle, i.e.,
$$
c^{t+s}=c^t+c^s\circ\phi^t
$$
holds for all $t,s\in\RR$. Moreover one has $\Ent(\omega_t|\omega)=-\omega(c^t)$ for all $t\in\RR$.
\eep

We denote by $Q_{\omega,t}$ the law of $c^t$ w.r.t. $\omega$ and set
\[
\fF_{\omega,t}(\alpha):=\int_\RR\e^{-\alpha s}\d Q_{\omega,t}(s)=\int_\cX \left(\frac{\d\omega}{\d\omega_{-t}}\right)^{-\alpha}\d\omega,\qquad(\alpha\in\CC).
\]

The next two assumptions will allow us to define the entropy production observable.

\assuming{Cl2}{ClTwo}{%
$c^t\in C(\cX)$ for all $t\in \RR$.
}

\assuming{Cl3}{ClThree}{%
The map $\RR\ni t\mapsto c^t\in C(\cX)$ is differentiable at $t=0$.
}

Note that \ClOne{} is the classical analog of Assumption~\RegOne, that
\ClTwo{} is the classical analog of Assumption~\RegTwo, and that~\ClThree{}
corresponds to the condition $V\in\Dom(\delta_\omega)$.

The {\sl entropy production} observable (or {\sl phase space contraction rate})
of $(X,\phi,\omega)$ is defined by
$$
\sigma=\frac{\d\ }{\d t}c^t\big|_{t=0}.
$$

\begin{theorem}\label{str-0-cl-1}\label{thm-qspc-ti-cl-1}
Assume that \ClOne--\ClThree{} hold. Then
\ben
\item
$$
c^t=\int_0^t \sigma_s \d s,
$$
and
$$
\Ent(\omega_t|\omega)=-\int_0^t \omega_s(\sigma)\d s.
$$
\item $\omega(\sigma)=0$.
\een

In the remaining statements we also assume that $(\cX,\phi,\omega)$ is {\rm TRI}
w.r.t.\;the time-reversal $\iota$.

\ben\setcounter{enumi}{2}
\item $c^t\circ \iota=c^{-t}$ for all $t\in\RR$, and $\sigma\circ\iota=-\sigma$.
\item For all $t\in\RR$ and $\alpha\in\CC$,
$$
\fF_{\omega, t}(\alpha)= {\overline{\fF_{\omega, t}(1-\bar \alpha)}}.
$$
\item
Let $\fr:\RR\to\RR$ be the reflection $\fr(s)=-s$ and $\bar Q_{\omega,t}=Q_{\omega,t}\circ\fr$.
Then the measures $Q_{\omega,t}$ and $\bar Q_{\omega,t}$ are mutually absolutely continuous and
\beq
\frac{\d\bar Q_{\omega,t}}{\d Q_{\omega,t}}(s)=\e^{-s}.
\label{str-2-cl-1}
\eeq
\een
\end{theorem}

For the proof we refer the reader to~\cite{Jaksic2011}.

We set
\[
P_{\omega, t}(B)=Q_{\omega, t}(tB).
\]
If the full/local LDP holds for $(P_{\omega,t})_{t>0}$, we will say that the
{\sl full/local classical Evans--Searles fluctuation theorem} holds for
$(\cX,\phi, \omega)$. The statement and the proof of Proposition~\ref{es-thm-1}
directly extend to the family $(P_{\omega,t})_{t>0}$; see~\cite{Cuneo2017}.
The relation~\eqref{str-2-cl-1} and the induced
symmetry~\eqref{quantum-es-1} of the LDP rate function for $(P_{\omega,t})_{t>0}$
constitute the {\sl classical Evans--Searles fluctuation relation.}

Until the end of this section we assume that~\ClOne--\ClThree{} hold.

Going beyond the reference state, for any state $\nu$ on $\cX$ we denote by
$Q_{\nu,t}$ the law of $c^t$ w.r.t.\;$\nu$. Let
\[
\fF_{\nu, t}(\alpha):=\int_\RR \e^{-\alpha s}\d Q_{\nu, t}(s), \qquad (\alpha\in\CC),
\]
and
$P_{\nu, t}(B)=Q_{\nu, t}(tB)$.
Since
\beq
\fF_{\nu, t}(\alpha)=\int_\cX \e^{-\alpha c^t(x)}\d\nu(x),
\label{eq:cl-laplace}
\eeq
if $\nu_n\to\nu$ weakly then $Q_{\nu_n,t}\to Q_{\nu,t}$ weakly.

Of particular interest is the case where $\nu$ is a NESS of $(\cX,\phi,\omega)$.
A NESS is a weak-limit point of the net
\[
\left\{\frac1T\int_0^T \omega_t\d t\,\bigg|\, T>0\right\}
\]
as $T\uparrow \infty$. The set of NESS is non-empty and any NESS is $\phi$-invariant.
Moreover, for any NESS $\omega_+$ one has
\[
\omega_+(\sigma)\geq 0.
\]
Until the end of this section we also assume the following classical analogue of~\NESS.

\assuming{Cl4}{ClFour}{%
The weak limit $\ds\lim_{t\to\infty}\omega_t=\omega_+$ exists so that, in particular, $\omega_+$
is the unique NESS of the system $(\cX, \phi, \omega)$.
}

If the full/local LDP holds for $(P_{\omega_+,t})_{t>0}$, we say that the
{\sl full/local classical Gallavotti--Cohen fluctuation theorem} holds for
$(\cX,\phi,\omega)$. If in addition the respective rate function $\II_+$ satisfies
\beq
\II_+(-s)= \II_+(s) +s
\label{classical-qc-1}
\eeq
on its domain,
we say that the {\sl classical Gallavotti--Cohen fluctuation relation} holds.
Unlike in the Evans--Searles case, \eqref{classical-qc-1} is not forced by
the LDP and time-reversal. The classical PREF and the related exchange of
limits argument that we discuss next is the only known general mechanism
that ensures its validity.

\begin{definition}
We say that the \textbf{full weak classical PREF} holds for $(\cX,\phi,\omega)$
if the families $(P_{\omega, t})_{t>0}$ and $(P_{\omega_+,t})_{t>0}$ both satisfy
a full LDP with the same rate function. The \textbf{local weak classical PREF}
holds if they satisfy a local LDP on the same interval $]-a, a[$, $a>0$, with
the same rate function.
\end{definition}

\begin{definition}
We say that $(\cX,\phi,\omega)$ satisfies the \textbf{strong classical PREF}
on an open interval $]\vartheta_-,\vartheta_+[$ containing $0$ if the limits
\beq
\begin{split}
\bF_{\omega}(\alpha)&:=\lim_{t\to\infty}\tfrac1t\log\fF_{\omega,t}(\alpha),\\[1mm]
\bF_{\omega_+}(\alpha)&:=\lim_{t\to\infty}\tfrac1t\log\fF_{\omega_+,t}(\alpha),
\end{split}
\label{thup-2}
\eeq
exist for all $\alpha\in]\vartheta_-,\vartheta_+[$, and define differentiable
functions on this interval satisfying
$$
\bF_{\omega}=\bF_{\omega_+}.
$$
\end{definition}

The remark after Definition~\ref{def-sqprf-1} applies to the strong classical PREF
as well, and we adopt the parallel terminology of local/full strong classical PREF.

The equality of the rate function in the full weak classical PREF is related
to an exchange of limits. Indeed, suppose that the  families $(P_{\omega,t})_{t>0}$ and
$(P_{\omega_+, t})_{t>0}$ satisfy  full LDP. Then by Varadhan's Lemma,
see e.g.~\cite[Theorem~4.3.1]{Dembo2000}, the limits
\[
\bF_{\omega}(\alpha)=\lim_{t\to\infty}\tfrac1t\log\fF_{\omega,t}(\alpha) \quad\text{and}\quad \bF_{\omega_+}(\alpha)=\lim_{t\to\infty}\tfrac1t\log\fF_{\omega_+,t}(\alpha)
\]
exist, and are obviously finite. The respective LDP rate functions satisfy
\beq
\II(s)=\sup_{\alpha\in\RR}\left(s\alpha-\bF_{\omega}(-\alpha)\right), \qquad
\II_+(s)=\sup_{\alpha\in\RR}\left(s\alpha-\bF_{\omega_+}(-\alpha)\right).
\label{rate-cl-1}
\eeq
By the basic property of the Legendre transform, $\II=\II_+$ iff $\bF_{\omega}=\bF_{\omega_+}$.
Note also that
\[
\fF_{\omega_T, t}(\alpha)=\int_\cX \e^{-\alpha c^{t}(x)}\d \omega_{T}(x)
=\int_\cX \e^{-\alpha c^{t}(x)}\e^{\int_0^T \sigma_{-s}(x) \d s}\d \omega(x),
\]
and so for all $\alpha \in \RR$,
\[
C_T^{-1}\fF_{\omega, t}(\alpha)\leq \fF_{\omega_T, t}(\alpha)\leq C_T\fF_{\omega, t}(\alpha),
\]
where
 $C_T=\e^{T\|\sigma\|_{\infty}}$. Thus, $\bF_{\omega}=\bF_{\omega_+}$ iff
\beq
\lim_{T\to\infty}\lim_{t\to\infty}\tfrac1t\log\fF_{\omega_T,t}(\alpha)
=\lim_{t\to\infty}\lim_{T\to\infty}\tfrac1t\log\fF_{\omega_T,t}(\alpha)
\label{cl-ex-lim-1}
\eeq
holds for all $\alpha\in\RR$.

In the case of the local weak PREF one cannot invoke Varadhan's lemma. However,
the strong PREF postulates the existence of the limits~\eqref{thup-2} and the
relations~\eqref{rate-cl-1} remain valid with suprema taken over
$\alpha\in]\vartheta_-,\vartheta_+[$. Again, $\II=\II_+$ iff
$\bF_{\omega}=\bF_{\omega_+} $ iff~\eqref{cl-ex-lim-1} holds.

The results of this section simplify in the case of discrete time dynamical
systems where $\phi=(\phi^t)_{t\in\ZZ}$. While  Assumptions~\ClOne{}
and~\ClTwo{} remain in force, \ClThree{} is dropped
and the entropy production observable is defined by
\[
\sigma:=\ell_{\omega_1|\omega}\circ \phi^1.
\]
Theorem~\ref{thm-qspc-ti-cl-1} extends directly to the discrete setting,
with simplified proofs and time integrals replaced with sums; see~\cite{Jaksic2011}.
The same applies to PREFs and the exchange of limits argument.

A celebrated example of this discrete time setting are Anosov diffeomorphisms of compact
Riemannian manifolds, see the seminal papers~\cite{Gallavotti1995b,Gallavotti1995c}. In the
context of the classical PREF this example has been discussed in~\cite{Jaksic2011,
Cuneo2017}, and the full strong classical PREF holds with
$\bF_{\omega_+}=\bF_{\omega}$ equal to the topological pressure of a
suitable Hölder continuous function on $\cX$. For Anosov diffeomorphisms the
map $\RR\ni\alpha\mapsto\bF_{\omega}(\alpha)$ is real-analytic.

\subsection{Quantum phase space contraction}
\label{sec-qpsc-1}

The extension of the notions and results of the previous section to quantum
dynamical systems $(\cO,\tau,\omega)$ relies on the non-commutative version of
the Radon--Nikodym derivative $\frac{\d\omega_t}{\d\omega}$ which, as suggested
by our notation, is related to the relative modular operators
$\Delta_{\omega_t|\omega}$ and the Araki--Connes family
$([D\omega_t:D\omega]_\alpha)_{t\in\RR,\alpha\in\i\RR}$. Besides the general
cocycle relation~\eqref{eq:gencocycle}, the latter also satisfies the following
multiplicative $\tau$-cocycle relation
\bep\label{start-qpsc-1}
For all $t,s\in\RR$ and $\alpha\in\i\RR$,
\beq
[D\omega_{t+s}:D\omega]_\alpha
=\tau^{-t}([D\omega_s:D\omega]_\alpha)[D\omega_t:D\omega]_\alpha.
\label{tom-vac-1}
\eeq
\eep
Since we lack a convenient reference, the proof of this result is given in
Section~\ref{sec-proof-start-qpsc-1}.

To proceed, we make two additional regularity assumptions.

\assuming{Qu1}{QuOne}{For all $t\in\RR$, the map
$\RR\ni\theta\mapsto[D\omega_t:D\omega]_{\i\theta}\in\cO$ is
differentiable at $\theta=0$.}

Let
\beq
\ell_{\omega_t|\omega}:= \frac1{\i}\frac{\d\ }{\d\theta}[D\omega_t:D\omega]_{\i\theta}\bigg|_{\theta=0},
\qquad
c^t:=\tau^t(\ell_{\omega_t|\omega}),
\label{eq:ctDef}
\eeq
and note that $\ell_{\omega_t|\omega}$, and consequently $c^t$, are self-adjoint elements of $\cO$.

\assuming{Qu2}{QuTwo}{The map $\RR\ni t\mapsto c^t\in\cO$ is differentiable at $t=0$.}
Let
$$
\sigma=\frac{\d\ }{\d t}\ell_{\omega_t|\omega}\bigg|_{t=0}= \frac{\d\ }{\d t}c^t\bigg|_{t=0}.
$$
The next result has been known to workers in the field for a long time;
see for example~\cite[Section 7.2]{Jaksic2012}. Since its proof has not appeared
in print, we will provide it in Section~\ref{sec-proof-thm-qpsc}.

\begin{theorem}\label{thm-qspc-ti-2} Suppose that \QuOne{} holds.
\ben
\item The family $(c_t)_{t\in\RR}$ is an additive $\tau$-cocycle: for any $s,t\in\RR$, one has
\beq
c^{t+s}=c^t+\tau^t(c^s).
\label{str-0-quantum}
\eeq

\item $\log\Delta_{\omega_t|\omega}=\log\Delta_\omega+\ell_{\omega_t|\omega}$
and\, $\Ent(\omega_t|\omega)=-\omega(c^t)$ hold for all $t\in\RR$.
\een

In the remaining statements we assume that~\QuTwo{} also holds.

\ben\setcounter{enumi}{2}
\item
\[
c^t=\int_0^t \sigma_s \d s,
\]
and
\beq {\rm Ent}(\omega_t|\omega)=-\int_0^t \omega_s(\sigma)\d s.
\label{ent-balance-2-1}
\eeq
 \item $\omega(\sigma)=0$.

\item If $(\cO, \tau, \omega)$ is {\rm TRI} with time-reversal map
$\Theta$, then $\Theta(c^t)=c^{-t}$ and $\Theta(\sigma)=-\sigma$.
\een
\end{theorem}

\noindent{\bf Remark 1.} $\sigma$ is called the {\sl entropy production observable} of
$(\cO,\tau,\omega)$ and~\eqref{ent-balance-2-1} is the {\sl entropy balance equation.}
The entropy balance equation has a long history in mathematical physics. It goes
back at least to~\cite{Pusz1978}\footnote{See the Remark on page
281. Another pioneering work on the subject is~\cite{Spohn1978b}.},
and was re-introduced independently in the literature several times since then;
see~\cite{Ojima1988,Ojima1989,Ojima1991, Jaksic2001a}. A basic consequence of the entropy
balance equation and the sign of relative entropy is that $\omega_+(\sigma)\geq
0$ for any NESS $\omega_+$.

\noindent{\bf Remark 2.} If $(\cO,\tau,\omega)$ is an open quantum system with
$V\in\Dom(\delta_\omega)$, then
\beq
\log\Delta_{\omega_t|\omega}=\log\Delta_\omega+\int_0^t\tau^{-s}(\delta_\omega(V))\d s,
\label{eq:logdelta}
\eeq
follows from~\cite[Lemma~2.4]{Benoist2024b}.
It is thus easy to check that Assumptions~\QuOne{} and~\QuTwo{} hold, and that
the entropy production observable is given by $\sigma=\delta_\omega(V)$.

\noindent{\bf Remark 3.} In further parallel with the classical case and
following \cite{Bochkov1977} one can take the spectral measure $Q_{\omega,
t}^{\rm naive}$ for $\omega$ and $c^t$ as a possible candidate for formulation
of a quantum Evans--Searls fluctuation theorem. However, for this  choice, which
is  in the literature  sometimes called the ``naive'' or ``direct'' quantization of
the classical $Q_{\omega, t}$ , in the TRI case the finite time Evans--Searles
fluctuation relation~\eqref{str-2-cl-1} {\sl fails}, see   {\sl e.g.}
\cite[Section 3.3]{Jaksic2010b}. It is precisely this failure that
motivated the early searches~\cite{Kurchan2000,Tasaki2000,Tasaki2003} for
alternative  candidates for a quantum Evans--Searles fluctuation theorem and
relation.

\medskip
Motivated by Proposition~\ref{start-qpsc-1} and Theorem~\ref{thm-qspc-ti-2}\footnote{Compare
them with Proposition~\ref{chain-cl} and Theorem~\ref{thm-qspc-ti-cl-1}. See also the next section.}
we consider the map
$$
\i \RR \ni \alpha \mapsto [D\omega_{-t}: D\omega]_{\alpha}\in \cO
$$
as a characterization of the {\em quantum phase space contraction} of $(\cO,\tau,\omega)$
at time $t$, and set\footnote{Compare with~\eqref{eq:cl-laplace}.}
\beq\label{qpscDef}
\fF_{\nu,t}^\qpsc(\alpha):=\nu([D\omega_{-t}:D\omega]_{\alpha})
\eeq
for $\nu\in\cS_\cO$. When $\nu=\omega$ this functional is linked to the 2TMEP and
the EAST protocols by the identities
\beq
\fF_{\omega,t}^\qpsc=\fF_{\omega,t}^\ttm=\fF_{\omega,t}^\anc.
\label{eq:reference-identical}
\eeq
These identities are broken as soon as $\omega$ is replaced by some other state $\nu$.

With the introduction of $\fF_{\nu,t}^\qpsc$ it appears natural to add to the strong quantum PREF of
Definition~\ref{def-sqprf-1} the requirement that also the limit
\[
\bF_{\omega_+}^\qpsc(\alpha):=\lim_{t\to\infty}\tfrac1t\log\fF_{\omega_+,t}^\qpsc(\alpha)
\]
exists for $\alpha\in]\vartheta_-,\vartheta_+[$, and that
$$
\bF^\ttm_{\omega}=\bF_{\omega_+}^\qpsc
$$
on this interval. We will refer to such PREF as strong + qpsc quantum PREF.\footnote{In the
sequel, whenever the meaning is clear within the context, we will simply write PREF for any
of its variants.}

\subsection{Comparison of the classical and quantum cases}
\label{sec-cl-q-comp}

We start with the algebraic description of the classical setting of
Section~\ref{sec-classical-1}. Let $\cO=C(\cX)$ and $\tau^t(f)=f\circ\phi^t$.
Obviously, $\cO$ is a commutative $C^\ast$-algebra and $\tau$ a
$C^\ast$-dynamics on $\cO$. If $\iota$ is a time-reversal of $(\cX,\phi)$,
$\Theta(f)=\bar f\circ\iota$ is the corresponding time-reversal of
$(\cO,\tau)$. The Banach space dual $\cO^\ast$ is identified with the
vector space of all Borel complex measures on $\cX$ equipped with total
variation norm.\;$\cS_\cO$ coincides with the set of Borel probability
measures on $\cX$ equipped with topology of weak convergence.

Fixing a reference state leads to a triple $(\cO,\tau,\omega)$. The
GNS representation $(\cH,\pi,\Omega)$ of $\cO$ associated to $\omega$ is
given by $\cH=L^2(\cX,\d\omega)$, $\pi(f)g=fg$, and $\Omega=\one$, the constant
function $\one(x)=1$ for $x\in \cX$. We then have
\[
\fM=\pi(\cO)^{\prime\prime}=L^\infty(\cX,\d\omega).
\]
The natural cone is ${\cH}_+=\{g\in\cH\mid g\geq0\}$ and the modular conjugation
is $J(g)=\bar g$. The state $\omega$ is automatically faithful, and a state $\nu$ is
$\omega$-normal iff $\nu\ll\omega$. In this case, the unique representative of
$\nu$ in $\cH_+$ is the vector
\[
\Omega_\nu=\left(\frac{\d\nu}{\d\omega}\right)^{1/2}.
\]
If $\nu\ll\rho$, then
\[
\Delta_{\nu|\rho}(g)=\frac{\d\nu}{\d\rho}g
\]
is the relative modular operator of the pair $(\nu,\rho)$. Note that
$\Delta_{\omega}=\one$ and $\log \Delta_\omega=0$. The triviality of
$\Delta_\omega$ is the crucial difference between the classical (commutative)
and the quantum (non-commutative) case.

The triviality of the classical $\Delta_\omega$ gives that in the classical
setting of Section~\ref{sec-classical-1}, Proposition~\ref{start-qpsc-1} reduces
to Proposition~\ref{chain-cl}, and similarly Theorem~\ref{thm-qspc-ti-2} reduces
to Theorem~\ref{thm-qspc-ti-cl-1}(1)-(2)-(3) while Proposition~\ref{gen-ttm-quant}
and Identities~\eqref{eq:reference-identical} give Theorem~\ref{thm-qspc-ti-cl-1}(4)-(5).
Thus, the quantum phase space construction
discussed in Section~\ref{sec-qpsc-1} is a natural non-commutative extension
of the classical theory described in Section~\ref{sec-classical-1}. But it is
not the only possible one: the functionals $\fF_{\nu,t}^\ttm$ and
$\fF_{\nu,t}^\anc$ are also non-commutative extensions of the classical $\fF_{\nu,t}$,
and in fact one can construct an entire host of mathematically natural
non-commutative extensions of $\fF_{\nu,t}$; see~\cite[Section~3.3]{Jaksic2010b}.
Some of these non-commutative extensions have direct quantum
mechanical interpretations, like $\fF_{\nu,t}^\ttm$ and $\fF_{\nu,t}^\anc$,
and some do not, or such interpretations are not known at the
moment.\footnote{For example, if $\nu\neq\omega$, then to the best of
our knowledge, the functional $\fF_{\nu,t}^\qpsc$ does not have a quantum
mechanical interpretation.}

The diversity of non-commutative/quantum theory of entropic fluctuations stems
from the rich KMS structure associated to $\omega$ via $\Delta_\omega$. Our
focus here is on the three non-commutative extensions described by
$\fF_{\nu,t}^\ttm$, $\fF_{\nu,t}^\anc$ and $\fF_{\nu,t}^\qpsc$, and their relation
to what one can reasonably call the quantum extension of the
classical Evans--Searles/Gallavotti--Cohen fluctuation theorem and the classical
PREF. Other routes are possible, and we will discuss some of them in a
forthcoming review paper. The emerging picture is that there is no unique theory
of entropy production and entropic fluctuations in quantum statistical mechanics
and that distinct approaches, with distinct physical interpretations, are linked
to the richness of the modular structure.

Returning to the results presented so far, we make the following remarks.

\noindent {\bf 1.} The rigidity result of Theorem~\ref{ver-fl-1} is due to the
ergodicity of the modular group(s)\footnote{See~\cite{ Benoist2023a} for a
discussion of this point.} and has no classical analog. To illustrate this,
suppose that in the classical setting of Section~\ref{sec-classical-1} one has
$Q_{\omega_T,t}=Q_{\omega,t}$ for all $t,T\ge0$. This implies $\omega(c^t\circ
\phi^T)=\omega(c^t)$ and the cocycle relation of Proposition~\ref{chain-cl}
gives
$$
\omega(c^{t+T})-\omega(c^T)=\omega(c^t).
$$
Dividing this identity with $t$ and taking $t\to0$ gives that, for all
$T>0$, $\omega(\sigma_T)=\omega(\sigma)=0$. This in turns implies that
$\omega(c^t)=0$ for all $t>0$, and the second assertion in
Proposition~\ref{chain-cl} gives that $\Ent(\omega_t|\omega)=0$, or equivalently
that $\omega_t=\omega$, for all $t>0$. Thus, in the classical setting the
stability result of Theorem~\ref{ver-fl-1}(1) is possible only if all entropic
quantities are identically equal to zero.\footnote{Needless to say, the
ergodicity assumption of Theorem~\ref{ver-fl-1}(1) is also never satisfied in
the classical case since the modular group is trivial.}

\noindent{\bf 2.} The weak classical PREF relies on two independent ingredients. The
first is the validity of the LDP for the families $(P_{\omega,t})_{t>0}$ and
$(P_{\omega_+,t})_{t>0}$, and the second is the equality of the respective rate
functions. In the case of the full weak classical PREF, the rate functions are
equal iff the exchange of limits~\eqref{cl-ex-lim-1} holds. The validity of this
exchange of limits is a strong ergodic type dynamical property of $(\cX,\phi,
\omega)$ which must be checked on a case by case basis and which typically
depends on the fine details of the dynamics. The strong classical PREF goes
further in the sense that it rests the validity of the LDP on the
Gärtner--Ellis theorem. Its naturalness partly stems from the
interpretation of $\bF_{\omega}$ and $\bF_{\omega_+}$ as spectral resonances
of classical transfer operators ; see~\cite{Jaksic2011} and
Section~\ref{sec-clas-tra} below.

The passage to the non-commutative/quantum theory comes with a number of
surprises that do not have classical analog. The first of them is the rigidity
result of Theorem~\ref{ver-fl-1} that essentially trivializes a very important
aspect of the PREF and gives that very generally the quantum Evans--Searles and
Gallavotti--Cohen fluctuation theorem are mathematically equivalent statement:
one holds iff the other holds. What is classically a fine model dependent
dynamical property of the system, in the quantum case follows (in a rather
strong form) from a modular ergodicity assumption that holds in paradigmatic
models of open quantum systems. The novel physical and mathematical
aspect of the strong quantum PREF concerns the ancilla part. The relation
\beq
\bF_{\omega}^\ttm=\bF_{\omega_+}^\anc
\label{sun-s-1}
\eeq
is a fine model dependent quantum dynamical property that can be seen as a non-trivial
counterpart of the classical PREF relation $\bF_{\omega}=\bF_{\omega_+}$. The
exchange of limits characterization of~\eqref{sun-s-1}, Relation~\eqref{qu-ex-lim-1},
parallels in its depth the exchange of
limits~\eqref{cl-ex-lim-1}. We again emphasize that under the assumptions of
Theorem~\ref{ver-fl-1}, the exchange of limits
$$
\lim_{T\to\infty}\lim_{t\to\infty}\tfrac1t\log\fF_{\omega_T,t}^\ttm(\alpha)=
\lim_{t\to\infty}\lim_{T\to\infty}\tfrac1t\log\fF_{\omega_T,t}^\ttm(\alpha)
$$
is a triviality.

\noindent{\bf 3.} The identities
\beq
\fF_{\omega,t}^\ttm=\fF_{\omega,t}^\anc=\fF_{\omega,t}^\qpsc
\label{iden-s-s}
\eeq
enrich the quantum Evans--Searles fluctuation theorem by providing additional physical
and mathematical interpretations of the 2TMEP. The identities~\eqref{iden-s-s} are broken
if $\omega$ is replaced by some other state $\nu$. However, while $\fF_{\nu,t}^\ttm$ and
$\fF_{\nu,t}^\anc$ have quantum mechanical interpretations for any state
$\nu$, such an interpretation is lacking for $\fF_{\nu,t}^\qpsc$ if $\nu\neq\omega$.
Thus, although the identities~\eqref{iden-s-s} are restored by the
strong + qpsc quantum PREF on the LDP scale for $\nu=\omega_+$ in the sense that
\beq
\bF_{\omega}^\ttm=\bF_{\omega_+}^\ttm=\bF_{\omega_+}^\anc=\bF_{\omega_+}^\qpsc,
\label{so-so-s-s}
\eeq
the last term $\bF_{\omega_+}^\qpsc$ lacks a physical interpretation. The first
equality in~\eqref{so-so-s-s} is immediate under the assumptions of
Theorem~\ref{ver-fl-1}, while the validity of
$\bF_{\omega}^\ttm=\bF_{\omega_+}^\qpsc$ is mathematically subtle but, we emphasize,
without physical interpretation. It is precisely the ancilla part
in~\eqref{so-so-s-s} that makes the strong + qpsc quantum PREF both
mathematically and physically as deep as its classical counterpart.

Due to its mathematical naturalness, we feel that the quantum phase space contraction
should be studied as an integral part of quantum theory of entropic fluctuations.

\noindent{\bf 4.} After this introduction, in the next two sections we will
describe quantum transfer operators and the spectral resonance theory of the
strong quantum PREF. These two sections are a non-commutative extension of the
classical theory described in~\cite{Jaksic2011}.They build on the quantum
transfer operator construction of NESS developed in~\cite{Jaksic2002a}; see
also~\cite{Jaksic2010b} for a pedagogical introduction to the topic. For
comparison purposes, we will briefly review the classical theory in
Section~\ref{sec-clas-tra}. The transfer operator connection supports the
naturalness of the strong PREFs.

\section{Quantum transfer operators}
\label{sec-quantum-tra}

We continue with the same setting: $(\cO,\tau,\omega)$ is a modular
$C^\ast$-quantum dynamical system whose modular structure is described in
Section~\ref{sec:repeat}. In particular, $(\cH,\pi,\Omega)$ denotes the
GNS representation of $\cO$ associated to $\omega$, and we write $A$ for $\pi(A)$
and $\tau$ for $\pi\circ\tau$ whenever the meaning is clear within the
context. $\fM=\pi(\cO)^{\prime\prime}$ and $\cL$ is the standard Liouvillean of
$\tau$ in the representation $\pi$.

Throughout this section Assumptions~\AnV{} and~\AnC{} play an important role.
Recall that, by Proposition~\ref{proja-1}, \AnV{} $\Rightarrow$~\AnC.

\subsection{One-parameter families of Liouvillians}

Throughout this section we assume that \AnC{} holds. For $\alpha\in\fS(\vartheta)$
we define the family $U_\alpha=(U_\alpha^t)_{t\in\RR}$ of maps on the GNS Hilbert space $\cH$ by
\beq
U_\alpha^t:=\e^{\i t \cL}J[D\omega_t: D\omega]_{\bar \alpha}J.
\label{tpar-L}
\eeq
The following proposition is proved in Section~\ref{sec-proof-qto-2}.

\bep\label{prop-qto-2}
\ben
\item For $\alpha\in\fS(\vartheta)$, $\RR\ni t\mapsto U_\alpha(t)$
is a $C_0$-group of bounded operators on $\cH$.
\item For $\alpha\in\i\RR$, $U_{\alpha}$ is a strongly continuous unitary group.
\item For all $A\in \fM$, $t\in\RR$ and $\alpha\in \fS(\vartheta)$
\[
U_\alpha^tAU_{-\bar \alpha}^{t\ast}=\tau^t(A).
\]
\item For all $t\in\RR$ and $\alpha\in \fS(\vartheta)$, $U_{\alpha}^{t\ast}=U_{-\bar\alpha}^{-t}$.
\een

\medskip
We denote by $\cL_{\alpha}$ the generator of\, $U_{\alpha}$ with the convention
$U_{\alpha}^t=\e^{\i t\cL_{\alpha}}$.

\ben\setcounter{enumi}{4}
\item Suppose that \AnV{} holds. Then, for all $\alpha\in\fS(\vartheta)$,
\beq
\begin{split}
\cL_{\alpha}&=\cL+JVJ-J\varsigma_\omega^{-\i\bar \alpha}(V)J\\[1mm]
            &=\cL_\free+V-J\varsigma_\omega^{-\i\bar \alpha}(V)J.
\end{split}
\label{sun-ho}
\eeq
\een
\eep

\noindent{\bf Remark 1.} It follows from~(1) that there exist constants
$M_{\alpha}$ and $m_{\alpha}$ such that for $t\in \RR$,
\begin{equation}\label{eq:Ualphabound}
\|U_{\alpha}^t\|\leq M_{\alpha}\e^{m_{\alpha}|t|}.
\end{equation}
If \AnV{} holds then, by Proposition~\ref{prop-qto-1},
one can take $M_{\alpha}=1$ and
\[
m_{\alpha}= \|\varsigma_\omega^{-\i \Re\,\alpha}(V)\| + \|V\|.
\]

\noindent{\bf Remark 2.} For later comparison with the classical case,
we note that the first formula in~\eqref{sun-ho} can be written as
$$
\cL_{\alpha}=\cL-\i\alpha\int_0^1J\varsigma_\omega^{-\i s\bar\alpha}(\sigma)J \d s.
$$
This formula is valid under more general conditions than \AnV.
We leave this topic to an interested reader.

\medskip
We will refer to the $U_\alpha^t$'s as the Quantum Transfer Operators, and to
$\cL_{\alpha}$ as the $\alpha$-Liouvillian of $(\cO,\tau,\omega)$. Obviously,
$\cL_{0}=\cL$, $\cL_{\alpha} $ is self-adjoint for $\alpha\in\i\RR$, and more
generally $\cL_{\alpha}^\ast=\cL_{-\bar\alpha}$. Writing
$\alpha=\frac12-\frac1p$ with $p\in[1,\infty]$, one can show that $\cL_\alpha$
isometrically implements the dynamics $\tau$ in the Araki--Masuda non-commutative
$L^p(\fM,\omega)$-space (see~\cite[Section~3.4]{Jaksic2010b}). For this reason
$\cL_\alpha$ is also called the $L^p$-Liouvillean.

If $\vartheta>1/2$ and $\alpha\in\i\RR$, we also set
\beq
\widehat\cL_\alpha=\cL_\free+\varsigma_\omega^{\i\alpha/2}(V)
-J\varsigma_\omega^{-\i(1-\bar\alpha)/2}(V)J.
\label{eq:hatLDef}
\eeq
The naturalness of $\widehat\cL_\alpha$ stems from its connection with
EAST which we describe in the next section. Note that
\beq
\widehat\cL_\alpha
=\Delta_\omega^{-\alpha/2}\cL_{1/2-\alpha}\Delta_\omega^{\alpha/2}.
\label{sun-tuluz}
\eeq

\subsection{Liouvillian representation of entropic functionals}

The representations described below, and proved in
Section~\ref{sec-proof-qto-1}, are the primary reasons for introducing the
$\alpha$-Liouvilleans in the context of this work.

\bep\label{prop-qto-4}
Suppose that \AnV{} holds for some $\vartheta>\frac12$. Then, for $\alpha\in\i\RR$
and $t,T\in\RR$:
\ben
\item
\[
\fF_{\omega,t}^\ttm(\alpha)=\langle\Omega,\e^{\i t\cL_{\shalf-\alpha}}\Omega\rangle.
\]
\item
\[
\fF_{\omega_T,t}^\qpsc(\alpha)
=\langle\Omega,\e^{\i T\cL_\shalf}\e^{\i t\cL_{\shalf-\alpha}}\Omega\rangle.
\]
\item
\[
\fF_{\omega_T,t}^\anc(\alpha)
=\langle\Omega,\e^{\i T\cL_\shalf}\e^{\i t\widehat\cL_\alpha}\Omega\rangle.
\]
\item The relation in part~(1) analytically extends to the strip $|\Re(\alpha-\frac12)|<\vartheta$,
the one in part~(2) extends to the strip $\frac{1}{2}-\vartheta<\Re(\alpha)<\vartheta$,
while the one in part~(3) extends to the strip $1-2\vartheta< \Re(\alpha)<2\vartheta$.
We note that the first and last strip contain the real interval $[0,1]$.
\een
\eep

\section{Spectral resonance theory of PREF}
\label{sec-spectral-PREF}

\subsection{Prologue}

The exponential asymptotics of a $C_0$-group is closely linked to the complex
resonances of its generator. We start with a discussion of two general results
elucidating this relation.

Let $U=(U^t)_{t\in\RR}$ be a $C_0$-group on a Hilbert space $\fH$ and $M,m>0$ be
such that
\beq
\|U^t\|\leq M\e^{m|t|}
\label{timed}
\eeq
for all $t\in\RR$. We denote by $L$ the generator of $U$ with the convention
$U^t=\e^{-\i tL}$. The estimate~\eqref{timed} gives that
$\sp(L)\subset\{z\in\CC\mid |\Im z|\leq m\}$. Moreover, for $\Im z >m$ one has
\[
(z-L)^{-1}=\frac1\i\int_0^\infty \e^{\i z t}\e^{-\i t L}\d t,
\]
which gives that
$$
\|(z-L)^{-1}\|\leq\frac{M}{|\Im z|-m}.
$$

\begin{proposition}\label{first-gen-sem-1}\label{jpr-1}
Let $\phi,\psi\in\fH$. Suppose that, for some $\mu>0$, the map
\[
z\mapsto f(z)=\bra\phi,(z-L)^{-1}\psi\ket,
\]
originally defined for $\Im z>m$, has a meromorphic continuation to the half-plane $\Im z>-\mu$
such that its only singularity in this half-plane is a pole of order $N$ at $\fr$.
Suppose also that for some $r>0$ and for $j=0,1$,\footnote{Here and in the following, $\partial$ denotes
the Wirtinger derivative of a function of a complex variable.}
\begin{equation}
\label{eq:resolventestimate}
\sup_{y>-\mu}\int\limits_{|x|>r}|(\partial^jf)(x+\i y)|^{2-j}\d x<\infty.
\end{equation}
Then, for all $\gamma$ satisfying $\max(-\Im\fr,0)<\gamma<\mu$, we have
\[
\bra\phi,\e^{-\i tL}\psi\ket=\e^{-\i t\fr}p(t)+O(\e^{-\gamma t})
\]
as $t\uparrow\infty$, where $p$ is a polynomial of degree $N-1$ such that $p(0)$ is
the residue of $f$ at $\fr$. In particular, if $N=1$ then $p$ is constant equal to this residue.
\end{proposition}

The number $\fr$ is called spectral resonance of the generator $L$.

The above proposition has a converse:

\begin{proposition}\label{second-gem-sem-1}\label{jpr-2}
Let $\phi,\psi\in\fH$ be such that for some $\fr\in\CC$, $\gamma>\max(0,-\Im\fr)$
and some polynomial $p$ of degree $N-1\ge0$ one has
\[
R(t)=\bra\phi,\e^{-\i tL}\psi\ket-\e^{-\i t\fr}p(t)=O(\e^{-\gamma t})
\]
as $t\uparrow \infty$. Then the function
\[
z\mapsto f(z)=\bra\phi,(z-L)^{-1}\psi\ket
\]
has a meromorphic continuation from the half-plane $\Im z>m$ to the half-plane
$\Im z>-\gamma$, and its only singularity there is a pole of order $N$ at $\fr$.
Moreover, for some $r>0$ and any\, $0<\mu<\gamma$ the estimate
\begin{equation}
\label{eq:resolventestimate2}
\sup_{y>-\mu}\int\limits_{|x|>r}|(\partial^jf)(x+\i y)|^{2-j}\d x <\infty
\end{equation}
holds for $j=0$. If $R$ is twice differentiable, with derivatives satisfying
$$
|R^{(k)}(t)|=O(\e^{-\gamma t}),\qquad(k=1,2),
$$
as $t\uparrow\infty$, then the estimate~\eqref{eq:resolventestimate2}
also holds for $j=1$.
\end{proposition}

Special cases of the above two results were formulated in~\cite[Section
5.4]{Jaksic2011}. We provide proofs in Sections~\ref{sec-int-1}
and~\ref{sec-int-2}. The results have obvious analogs in the case of convention
$U^t=\e^{\i tL}$, where  the meromorphic continuation of the resolvent is taken
from the half-plane $\Im z<-m$ to $\Im z<\mu$ for some $\mu>0$. Alternatively,
with this convention one can simply apply Propositions~\ref{jpr-1}
and~\ref{jpr-2} to $-L$, and this is the way we will proceed.

Proposition~\ref{prop-qto-4} and~\ref{first-gen-sem-1} offer a spectral
resonance route to the verification of the strong PREF. The route is however
somewhat indirect, and we start by reviewing the spectral theory of NESS
developed in~\cite{Jaksic2002a}.

\subsection{Spectral theory of NESS}
\label{sec-spectNESS}

Throughout this section we assume that Assumption \AnV{} holds with
$\vartheta>1/2$. Central to the spectral theory of NESS is the
$\frac12$-Liouvillean $\cL_{\shalf}$. Note that
\[
\cL_\shalf=\cL_\free+V-J\Delta^{1/2}V\Delta^{-1/2}J,
\]
hence $\cL_\shalf\Omega=0$. Proposition~\ref{prop-qto-2}(3)-(4) give that, for
$A\in\cO$,
\beq
\omega\circ\tau^t(A)
=\langle\Omega,\e^{\i t\cL_\shalf} A\e^{-\i t\cL_\shalf}\Omega\rangle
=\langle\Omega,\e^{\i t\cL_\shalf}A\Omega\rangle.
\label{hhis}
\eeq
The next assumption sets an abstract spectral deformation scheme that leads
to the spectral theory of NESS.

\assuming{Deform1}{DeformOne}{%
There exists a bounded operator $D\geq0$ on $\cH$ such that $\Ran D$ is dense
in $\cH$,  $D\Omega=\Omega$, and that the following holds:
\begin{enumerate}[(a)]
\item The set $\cO_D=\left\{A\in\cO\mid A\Omega\in\Dom(D^{-1})\right\}$ is dense in $\cO$.
\item The map
\beq
z\mapsto F(z)=D(z+\cL_\shalf)^{-1}D\in \cB(\cH),
\label{st-d}
\eeq
originally defined for $\Im z>m_\shalf$, has a meromorphic continuation to a half-plane
$\Im z>-\mu$ for some $\mu>0$ such that its only singularity in this half-plane is a
simple pole at zero with residue $\cR_\shalf$.
\item For some $r>0$ and $j=0,1$,
\[
\sup_{y>-\mu}\int\limits_{|x|>r}\|\partial^j F(x+\i y)\|^{2-j}\d x <\infty.
\]
\end{enumerate}
}

Note that since $\Omega\in\ker\cL_{1/2}$, the singularity at $0$ of the map~\eqref{st-d}
is forced by the relation
$$
\langle \Omega, (z+\cL_\shalf)^{-1}\Omega\rangle=\frac1z.
$$
An immediate consequence of Proposition~\ref{first-gen-sem-1} and~\eqref{hhis} is
the following result of~\cite{Jaksic2002a}:

\begin{theorem}\label{thm-ness}
Suppose that \DeformOne{} holds. Then the limit
\beq
\omega_+(A)=\lim_{t\to\infty}\omega\circ\tau^t(A)
\label{m-win0}
\eeq
exists for all $A\in \cO$. For $A\in\cO_D$ the convergence is exponentially fast, {\sl i.e.,}
\beq
|\omega_+(A)-\omega\circ\tau^t(A)|=O(\e^{-\gamma t})
\label{m-win1}
\eeq
as $t\uparrow\infty$ for any $0<\gamma<\mu$, and
\beq
\omega_+(A)=\langle\Omega,\cR_\shalf D^{-1}A\Omega\rangle.
\label{m-win2}
\eeq
\end{theorem}
\proof Note that by~\DeformOne(b+c) Proposition~\ref{first-gen-sem-1} holds for all
$\phi,\psi\in\Ran D$ with $L=-\cL_\shalf$, $\fr=0$, and constant polynomial
\[
p=\langle D^{-1}\phi,\cR_\shalf D^{-1}\psi\rangle.
\]
By~\DeformOne(a), $A\Omega=DD^{-1}A\Omega\in\Ran D$ for $A\in\cO_D$, and so
\eqref{hhis} and Proposition~\ref{first-gen-sem-1} yield~\eqref{m-win1}
and~\eqref{m-win2}. Since $\cO_D$ is dense in $\cO$,
\eqref{m-win1} $\Rightarrow$~\eqref{m-win0}. \hfill\qed

\subsection{Spectral theory of PREF}
\label{sec-spectPREF}

We continue to assume that \AnV{} holds with $\vartheta>1/2$.
For $\zeta>0$ let
\[
B(\vartheta,\zeta):=\big\{z\in\CC\mid |\Re z|<\vartheta,\text{ and }|\Im z|<\zeta\big\}.
\]
We strengthen \DeformOne{} to

\assuming{Deform2}{DeformTwo}{%
Assumption \DeformOne{} holds and there exists $\zeta>0$ such that:
\begin{enumerate}[(a)]
\item For all $\alpha\in B(\vartheta,\zeta)$ and all $t\in\RR$,
$[D\omega_t:D\omega]_{\alpha}\in\cO_D$.
\item For all $\alpha\in B(\vartheta,\zeta)$ and all $t\in\RR$,
$[D\omega_{t}:D\omega]^\ast_{\frac{\bar\alpha}{2}}
[D\omega_{t}:D\omega]_{\frac{\alpha}{2}}\in\cO_D$.
\item For all $\alpha\in B(\vartheta,\zeta)$ there exists $\mu_\alpha>0$ such that the map
$$
z\mapsto F_\alpha(z)=D(z+\cL_{\alpha})^{-1}D\in \cB(\cH),
$$
originally defined on the half-plane $\Im z>m_\alpha$ (see Equation~\eqref{eq:Ualphabound}),
has a meromorphic continuation to the half-plane $\Im z>-\mu_\alpha$ whose only
singularity in this half-plane is a pole at $\cE(\alpha)$ with residue $\cR_\alpha$.
\item For all $\alpha\in B(\vartheta,\zeta)$ there exists $r_\alpha>0$ such that,
for $j=0,1$ and all $\phi,\psi\in\cH$,
\[
\sup_{y>-\mu_\alpha}\int\limits_{|x|>r_\alpha}
|\langle\phi,(\partial^jF_\alpha)(x+\i y)\psi\rangle|^{2-j}\d x<\infty.
\]
\item $\cR_\shalf^\ast\Omega\in\Dom(D^{-2})$.
\item For all $\alpha\in B(\vartheta,\zeta)$,
$D^{-1}\cR_\shalf^\ast\Omega\in\Dom(\Delta^{\alpha/2})$ and
$\Delta^{\alpha/2}D^{-1}\cR_\shalf^\ast\Omega\in\Dom(D^{-1})$.
\item For all $\alpha\in B(\vartheta,\zeta)$,
$$
\langle\Omega,\cR_\alpha\Omega\rangle\neq0,\quad
\langle D^{-1}\cR_\shalf^\ast\Omega,\cR_\alpha\Omega\rangle\neq0,\quad
\langle D^{-1}\Delta^{-\bar\alpha/2}D^{-1}\cR_\shalf^\ast\Omega,\cR_\alpha\Omega\rangle \neq0.
$$
\end{enumerate}
}

\noindent{\bf Remark.} Note that~(f) with $\alpha=0$ reduces to~(e). We have separated the
two assumptions because of their roles in the proof, and because of possible
alternative axiomatic schemes in which~(f) is bypassed; see Remark~3 at the end of this section.

\begin{theorem}\label{main-spect-def}
Under Assumption~\DeformTwo{} the following hold:
\ben
\item $\cE(\alpha)\in\i\RR$ for $\alpha\in]-\vartheta,\vartheta[$.
\item The limits
\beq
\begin{split}
\bF^\ttm_{\omega}(\alpha)
&=\lim_{t\to\infty}\tfrac1t\log\fF_{\omega,t}^\ttm(\alpha),\\[1mm]
\bF^\anc_{\omega_+}(\alpha)
&=\lim_{t\to\infty}\tfrac1t\log\fF_{\omega_+,t}^\anc(\alpha),\\[1mm]
\bF^\qpsc_{\omega_+}(\alpha)
&=\lim_{t\to\infty}\tfrac1t\log\fF^\qpsc_{\omega_+,t}(\alpha),
\end{split}
\label{thup-thup-1}
\eeq
exist for $\alpha\in]-\vartheta+\frac12,\vartheta[$,
and for $\alpha$ in this interval,
$$
\bF_{\omega}^\ttm(\alpha)=\bF_{\omega_+}^\anc(\alpha)=\bF_{\omega_+}^\qpsc(\alpha)=
-\i\cE\left(\tfrac12-\alpha\right).
$$
\item If in addition the assumptions of Theorem~\ref{ver-fl-1} are satisfied,
\footnote{Recall also the remark after this theorem; for Part~(3) we require that
$\omega_{+\sS}>0$.} then also
\beq
\bF^\ttm_{\omega_+}(\alpha)=\lim_{t\to\infty}\tfrac1t\log\fF_{\omega_+,t}^\ttm(\alpha)
\label{ssd-sd}
\eeq
exists for $\alpha\in]-\vartheta +\frac12,\vartheta[$ and
\beq
\bF_{\omega_+}^\ttm(\alpha)=-\i\cE\left(\tfrac12-\alpha\right).
\label{ssd-sd-1}
\eeq
\een
\end{theorem}

{\bf Remark.} The first limit in~\eqref{thup-thup-1} actually exists for
$\alpha\in]-\vartheta +\frac12,\vartheta+\frac12[$.

The proof gives that~\eqref{thup-thup-1} holds for all $\alpha\in\CC$
such that $-\vartheta+\frac12<\Re\alpha<\vartheta$ and $|\Im\alpha|< \zeta$.

We emphasize that~\eqref{ssd-sd} and~\eqref{ssd-sd-1} follow from
Theorem~\ref{ver-fl-1} and the induced equality $\bF_{\omega}^\ttm=\bF_{\omega_+}^\ttm$,
and do not require a proof.

To establish strong + qpsc PREF, Theorem~\ref{main-spect-def} needs to be
complemented with a study of the regularity of $\cE$ on $(-\vartheta, \vartheta)$,
and indeed the axiomatic scheme \DeformTwo{} can be further strengthened to also yield
the analyticity of $\cE$ on $B(\vartheta,\zeta)$.

\assuming{Deform3}{DeformThree}{%
Assumption \DeformTwo{} holds for some $\zeta>0$ such that any $\alpha_0\in B(\vartheta,\zeta)$
has a neighborhood $U\subset B(\vartheta,\zeta)$ with the following properties:
\begin{enumerate}[(a)]
\item For $\alpha\in U$, Conditions~\DeformTwo(c+d) hold with constants $r_\alpha=r_U$ and $\mu_\alpha=\mu_U$
which do not depend on $\alpha$.
\item
\[
\max_{j\in\{0,1\}}\sup_{\alpha\in U}\sup_{y>-\mu_U}\int\limits_{|x|>r_U}
|\langle\phi,(\partial^jF_\alpha)(x+\i y)\psi\rangle|^{2-j}\d x<\infty.
\]
\item For any $\epsilon>0$ there exists $C_\epsilon>0$ such that
\[
\sup_{\alpha\in U}\sup_{{\Im z >-\mu}\atop{|z-\cE(\alpha)|>\epsilon}}\|F_\alpha(z)\|<C_\epsilon.
\]
\item $\inf_{\alpha\in U}|\langle\Omega,\cR_\alpha\Omega\rangle|>0$.
\end{enumerate}
}
\begin{theorem}\label{ss-tuluz}
Suppose that \DeformThree{} holds. Then, in addition to the conclusions
of Theorem~\ref{main-spect-def}, the map
\[
B(\vartheta, \zeta)\ni \alpha \mapsto \cE(\alpha)
\]
is analytic. In particular strong + qpsc PREF holds on $]\frac12-\vartheta,\vartheta[$.
\end{theorem}

 We finish with four remarks.

\noindent {\bf Remark 1.} As already mentioned, the spectral resonance approach
to the PREF described in this section stems from Propositions~\ref{prop-qto-4}
and~\ref{first-gen-sem-1}, and the axiomatic schemes~\DeformOne{}-\DeformThree{}
are natural in view of the spectral deformation techniques developed in the
theory of Schrödinger
operators~\cite{Aguilar1971,Balslev1971,Simon1973,Hunziker1983,Hunziker1990},
see also~\cite[Chapter~8]{Cycon1987}, \cite[Sections~XII.6
and~XII.10]{Reed1978}. The schemes can be simplified in particular settings
where additional structural information is available, like the Spin--Fermion
model, or adjusted to the settings where they cannot be directly applied, like
the Spin--Boson model where $V$ is unbounded.

\noindent{\bf Remark 2.} Since $\cL_\alpha$ is an analytic family of type A, see
{\sl e.g.}~\cite{Kato1966,Reed1978}, for $\alpha\in\fS(\vartheta)$, it is
natural to expect that in concrete models for which~\DeformTwo{} holds, the
proof also gives that $\cE$ is analytic on $B(\vartheta,\zeta)$. This is indeed
the case for Spin--Fermion systems which we study in a continuation of this
work~\cite{Benoist2024c}.

\noindent{\bf Remark 3.} Proposition~\ref{prop-qto-4} (3) offers a parallel
spectral resonance approach to the ancilla part of PREF. It starts by
formulating a variant of~\DeformTwo{} for the family $\widehat {\cL}_\alpha$:

\assuming{Deform2A}{DeformTwoA}{%
Assumption {\bf (Deform1)} holds and there exists $\zeta >0$ such that:
\begin{enumerate}[(a)]
\item For all $\alpha \in B(\vartheta,\zeta)$ and all $t$,
\[
[D\omega_t: D\omega]^\ast_{\frac{\bar \alpha}{2}}
[D\omega_t: D\omega]_{\frac{\alpha}{2}}\in \cO_D.
\]
\item For all $\alpha\in B(\vartheta,\zeta)$ there exists $\mu_\alpha >0$ such that the function
\[
z\mapsto \widehat{F}_\alpha(z)=D( z +\widehat\cL_\alpha)^{-1} D\in \cB(\cH),
\]
originally defined for $\Im z>\widehat m_\alpha$, has a meromorphic continuation to the
half-plane $\Im z>-\mu_\alpha$ such that its only singularity in this half-plane
is a pole at $\widehat\cE(\alpha)$ with residue $\widehat\cR_\alpha$.

\item For all $\alpha\in B(\vartheta,\zeta)$ there exists $r_\alpha>0$ such that for
$j=0,1$ and all $\phi,\psi\in\cH$,
\[
\sup_{y>-\mu_\alpha}\int\limits_{|x|>r_\alpha}|
\langle \phi,(\partial^j\widehat{F}_\alpha)(x+\i y)\psi\rangle |^{2-j}\d x<\infty.
\]

\item $\widehat\cR_{1/2}^\ast\Omega\in\Dom(D^{-2})$.

\item $\langle D^{-1}\widehat\cR_{1/2}^\ast\Omega,\widehat\cR_\alpha\Omega\rangle\not=0$.
\end{enumerate}
}

Proposition~\ref{prop-qto-4} and~\DeformTwoA{} yield that for $\alpha\in]-\vartheta, \vartheta[$
\[
\lim_{t\to\infty}\tfrac1t\log\fF_{\omega_+,t}^\anc(\alpha)=-\i\widehat\cE(\alpha),
\]
with exactly the same proof as in Theorem~\ref{main-spect-def}. One then uses~\eqref{sun-tuluz}
to show that $\widehat\cE(\alpha)=\cE(\frac12-\alpha)$. This eliminates the need to
verify~\DeformTwo(f) and the third relation in~\DeformTwo(g), and potentially offers a
technically simpler route for the verification of the ancilla part of the PREF.
This again will be the case for the Spin--Fermion model.

\noindent {\bf Remark 4.} In summary, \DeformOne{}--\DeformThree{} should be viewed as an
adjustable frame for the study of PREF in concrete models. As one may expect given the information
they yield, it is technically involved to verify these axiomatic schemes in physically relevant
models. There is a strong parallel between the proposed spectral resonance approach to PREF
and the use of transfer operators in classical dynamical systems, and we will comment more
on this point in the next two sections.


\subsection{Classical dynamical systems}
\label{sec-clas-tra}

The spectral resonance approach and axiomatic scheme proposed in the previous
sections can actually also be used to derive the strong classical PREF.
We return to the setting of Section~\ref{sec-classical-1} and assume that
Assumptions~\ClOne, \ClTwo{} and~\ClThree{} hold. We will work in the
classical algebraic setting described in Section~\ref{sec-cl-q-comp} and
slightly adjust the presentation of the classical transfer operator theory
of~\cite{Jaksic2011} for easier comparison with general Liouvilleans introduced
here.

By Assumption~\ClOne{} the Koopman map $U^t_\koop f = f_t$ extends from
$C(\cX)$ to a $C_0$-group of bounded operators on the GNS Hilbert space
$\cH=L^2(\cX,\d\omega)$ which we denote by the same symbol. The standard
Liouvillean $\cL$ of the classical  system $(\cO,\tau,\omega)$ is the
generator of the $C_0$-group
\[
U_\liouv^tf=U_\koop^t\pi\left(\frac{\d\omega_t}{\d\omega}\right)^{-\frac12}f
=\pi(\e^{-\frac12\int_0^t\sigma_s\d s})U_\koop^tf,
\]
with the usual convention $U_\liouv^t=\e^{\i t\cL}$.
One easily checks that $U_\liouv$ is a unitary group on $\cH$ that
preserves the positive cone $\cH_+$ and implements the classical flow,
\[
U_\liouv^t\pi(f)U_\liouv^{t\ast}=\pi(f_t).
\]
For $\alpha \in \CC$ we define maps $U_\alpha^t:\cH\to\cH$ by
\[
U_\alpha^t f:=\e^{\i t\cL}J\pi\left(\frac{\d\omega_t}{\d\omega}\right)^{\bar\alpha} J f,
\]
which is of course the commutative case of~\eqref{tpar-L}. Note that
\[
U_\alpha^tf=\pi\left(\e^{(\alpha-1/2)\int_0^t \sigma_s \d s}\right) U_\koop^tf.
\]
The following is the classical analogue of Proposition~\ref{prop-qto-2}.
\bep
\ben
\item $U_{\alpha}:=(U_\alpha^t)_{t\in\RR}$ is a $C_0$-group of bounded
operators on $\cH$. If $\alpha \in \i\RR$, then $U_{\alpha}$ is a unitary group.
\item For all $f\in \cO$, $t\in\RR$ and $\alpha\in\CC$,
\[
U_\alpha^t\pi(f)U_{-\bar\alpha}^{t\ast}=\pi(f_t).
\]
\item For all $t\in\RR$ and $\alpha\in\CC$, $U_\alpha^{t\ast}=U_{-\bar\alpha}^{-t}$.
\item $U_0= U_\liouv$ and $U_\shalf= U_\koop$.
\een
\eep
We denote by $\cL_\alpha$ the generator of $U_\alpha$ with the
convention $U_\alpha^t=\e^{\i t\cL_\alpha}$. We will refer to
$\cL_\alpha$ as the classical $\alpha$-Liouvillean. Obviously,
\[
\cL_\alpha=\cL+\i\alpha\pi(\sigma).
\]
For any $\alpha \in \CC$ we have the following representations:
\beq\label{eq:classicalqto}
\begin{split}
\fF_{\omega,t}(\alpha)&=\int_{\cX}\e^{-\alpha\int_0^t\sigma_s\d s}\d\omega
=\langle\Omega,\e^{\i t\cL_{\shalf-\alpha}}\Omega\rangle,\\[2mm]
\fF_{\omega_T,t}(\alpha)&=\int_{\cX}\e^{-\alpha\int_0^t\sigma_s\d s}\d\omega_T
=\langle\Omega,\e^{\i T\cL_{1/2}}\e^{\i t\cL_{\shalf-\alpha}}\Omega\rangle.
\end{split}
\eeq

It follows that the axiomatic spectral approach to NESS and PREF,
described in Section~\ref{sec-spectNESS} and~\ref{sec-spectPREF},
applies directly to the classical setting. In particular the
representation~\eqref{eq:classicalqto}, together with a spectral
resonance analysis of the classical $\alpha$-Liouvilleans,
gives a route to the proof of the exchange of limit~\eqref{cl-ex-lim-1},
hence of the strong classical PREF.

The results of this section are easily adapted to the discrete time dynamical system;
see~\cite[Sections 10 and 11]{Jaksic2011} where the reader can also find non-trivial
examples to which this spectral approach applies.

\subsection{Remarks}
\label{sec-remarks-tuluz}

\noindent{\bf Remark 1.} In the context of finite quantum systems, the
$\alpha$-Liouvilleans were introduced in~\cite[Section 3.4]{Jaksic2010b}, where
also Part~(1) of Proposition~\ref{prop-qto-4} is established. Moreover, in the
same reference $\alpha$-Liouvilleans are linked to Araki--Masuda non-commutative
$L_p$ spaces~\cite{Araki1982}. Although we believe that this link to modular theory
is central to the understanding of the results of this work, for
reasons of space we postpone its further discussion.

\noindent{\bf Remark 2.} An early discussion of $\fF_{\omega, t}^\ttm$ can be
found in~\cite{DeRoeck2009} in the context of the Spin--Boson model.
In the same reference it is proven that for some $\vartheta >0$
and $\alpha\in(-\vartheta,\vartheta)$ the limit
\beq
\lim_{t\to\infty}\tfrac1t\log\fF_{\omega,t}^\ttm(\alpha)
\label{dr-lim-tuluz}
\eeq
exists and is real-analytic on $(-\vartheta,\vartheta)$. The proof is based on a
dynamical perturbative variant of complex spectral deformation technique applied
to a different class of transfer operators. Although we consider this work
pioneering, the transfer operators used there are not algebraically natural and
do not respect the fabric of the modular structure of quantum entropic
fluctuations which is, in our opinion, central for the topic. In the context of
the Spin--Fermion model, the
limit~\eqref{dr-lim-tuluz} has been informally discussed at the end
of~\cite[Section 5.4]{Jaksic2010b}. This discussion is in the spirit
of~\cite{Jaksic2002a} and the axiomatic scheme~\DeformOne--\DeformTwo.

\noindent{\bf Remark 3.} Returning to the classical dynamical system setting, in
our opinion the closest parallel to our spectral resonance theory described in
Sections~\ref{sec-quantum-tra} and~\ref{sec-spectral-PREF} is Ruelle’s abstract
presentation of statistical mechanics on Smale spaces~\cite[Section~7]{Ruelle2004}.
The parallel runs deep and concerns motivation/axioms
selection/role of transfer operators and even certain technical aspects of the
proof. The strong abstract chaoticity assumptions of Ruelle are matched by the
strong dynamical assumptions inherent to~\DeformOne--\DeformThree. Ruelle’s
axiomatization is motivated by the example of Anosov diffeomorphisms, and ours
by the Spin--Boson and Spin--Fermion models. The
hyperbolic structure of Anosov diffeomorphisms corresponds to strong dispersion
of free quantum reservoirs and the resulting dynamical Fermi Golden Rule.%
\footnote{ To learn more about these points, an interested reader can
compare the results/presentation of~\cite{Jaksic2011} and of this work.}

\noindent{\bf Remark 4.}  This  paper builds on concepts, techniques and results
of modern non-equilibrium quantum statistical mechanics that were
developed in the last twenty-five years starting with the articles~\cite{Ruelle2000,
Jaksic2001a,Ruelle2001}. The history of the subject is quite intricate.
For example, the relation between the works~\cite{Kurchan2000,Tasaki2000}
and~\cite{Tasaki2003}, which is central for the subject and was known to the experts
since 2003, did not appear in print until the work~\cite{Jaksic2010b}.%
\footnote{The  works~\cite{Kurchan2000, Tasaki2000} remain unpublished.}

\section{Proofs }
\label{sec-proofs-sr}
\subsection{Proof of Proposition~\ref{ancilla-start-1}}
\label{sec-proof-ancilla-1}

We identify $\widehat\cO=\cO\otimes\Mat_2(\CC)$ with the $C^\ast$-algebra
$\Mat_2(\cO)$ of all $2\times 2$ matrices with entries in $\cO$. Recall that
$\cL_\free$ is the standard Liouvillean of the free dynamics $\tau_\free$. Then,
$(\cL_\free+V)\otimes\Id+\widehat{W}_\alpha$ is the semi-standard Liouvillean of
$\widehat{\tau}_\alpha$, and hence
\[
U_\alpha^t=\e^{\i t((\cL_\free+V)\otimes\Id+\widehat{W}_\alpha)}=
\begin{bmatrix}\e^{\i t(\cL_\free+V+W_\alpha)} &0\\[1mm]
0 &\e^{\i t(\cL_\free+V+W_{-\alpha})}\end{bmatrix}
\]
satisfies
\[
\widehat\tau_\alpha^t(A)=U_\alpha^tAU^{t\ast}_\alpha
\]
for $A\in\widehat\cO$. An elementary calculation shows that, for $A\in\Mat_2(\CC)$ and
$\widehat{\nu}=\nu\otimes\rho$,
$$
\widehat{\nu}\circ\widehat{\tau}_\alpha^t(\one\otimes A)=
\sum_{r,s\in\{\pm\}}\rho_{rs}A_{sr}
\nu\left(\e^{\i t(\cL_\free+V+W_{s\alpha})}\e^{-\i t(\cL_\free+V+W_{r\alpha})}\right),
$$
and so it suffices to show that
\[\nu\left(\e^{\i t(\cL_\free+V+W_{-\alpha})}\e^{-\i t(\cL_\free+V+W_{\alpha})}\right)=\fF_{\nu,t}^\anc(\alpha).\]
Writing
$$
\e^{\i t(\cL_\free+V+W_{-\alpha})}
=\Delta_\omega^{-\alpha/2}\e^{\i t(\cL_\free+V)}\Delta_\omega^{\alpha/2}
=\left[\Delta_\omega^{-\alpha/2}\e^{\i t(\cL_\free+V)}\Delta_\omega^{\alpha/2}
\e^{-\i t(\cL_\free+V)}\right]\e^{\i t(\cL_\free+V)},
$$
we note that, since $\cL_\free$ and $\cL_\omega=\log\Delta_\omega$ commute
and $\sigma=\delta_\omega(V)=\i[\cL_\omega,V]\in\cO$,
$$
\frac{\d\ }{\d t}\e^{\i t(\cL_\free+V)}\cL_\omega\e^{-\i t(\cL_\free+V)}
=-\e^{\i t(\cL_\free+V)}\i[\cL_\omega,V]\e^{-\i t(\cL_\free+V)}
=-\tau^t(\sigma).
$$
Thus
\beq
\e^{\i t(\cL_\free+V+W_{-\alpha})}=\left[\Delta_\omega^{-\alpha/2}
\e^{\alpha(\cL_\omega-\int_0^t\tau^s(\sigma)\d s)/2}\right]\e^{\i t(\cL_\free+V)},
\label{eq:proofprop2.2}
\eeq
and since $\bar\alpha=-\alpha$, invoking Theorem~\ref{thm-qspc-ti-2}(2) and
recalling Relation~\eqref{eq:logdelta}, \eqref{eq:proofprop2.2} gives
\[
\e^{\i t(\cL_\free+V+W_{-\alpha})}
=[D\omega_{-t}: D\omega]^\ast_{\frac{\bar\alpha}{2}}\e^{\i t (\cL_\free +V)}.
\]
Reversing the sign of $\alpha$ we get
\beq
\e^{-\i t(\cL_\free +V +W_\alpha)}
=\left([D\omega_{-t}:D\omega]^\ast_{\frac{\alpha}{2}}\e^{\i t(\cL_\free+V)}\right)^\ast
=\e^{-\i t(\cL_\free+V)}[D\omega_{-t}:D\omega]_{\frac{ \alpha}{2}},
\label{cb-p}
\eeq
and the result follows.

\subsection{Proof of Proposition~\ref{prop-qto-1}}
\label{sec-proof-gto-1}

The relation~\eqref{cb-p} gives that, for $\alpha\in\i\RR$,
\beq
[D\omega_t:D\omega]_\alpha=\e^{-\i t(\cL_\free+V)}\e^{\i t(\cL_\free +V +W_{2 \alpha})},
\label{an-sa}
\eeq
where $W_{2\alpha}=\varsigma_\omega^{-\i\alpha}(V)-V$. Writing the Dyson expansion of the
right-hand side of the previous relation, we get
\beq
[D\omega_t:D\omega]_\alpha=
\one+\sum_{n\geq1}(\i t)^n\int\limits_{0\leq\theta_1\leq\cdots\le\theta_n\leq 1}
\tau^{-t\theta_n}(W_{2\alpha})\cdots\tau^{-t\theta_1}(W_{2\alpha})\d\theta_1\cdots\d\theta_n.
\label{mar-exp}
\eeq
Assumption~\AnV{} gives that the function
\[
\i\RR\ni\alpha\mapsto W_{2\alpha}\in\cO
\]
has an analytic extension to the strip $\fS(\vartheta)$. It follows from the right-hand side
of Relation~\eqref{mar-exp} that the same holds for the map
$$
\i\RR\ni\alpha\mapsto[D\omega_t: D\omega]_\alpha\in\cO.
$$
The expansion~\eqref{mar-exp} also yields the estimate~\eqref{est-sa}.

\subsection{Proof of Proposition~\ref{te-pa-1}}
\label{sec-proof-te-pa}

The unitarity of the Araki--Connes cocycle yields that, for $\alpha\in\i\RR$ and $t\in\RR$,
$$
[D\omega_t:D\omega]_{-\bar\alpha}^\ast[D\omega_t:D\omega]_\alpha
=[D\omega_t:D\omega]_\alpha^\ast[D\omega_t:D\omega]_\alpha=\one.
$$
Under Assumption~\AnV{}, Proposition~\ref{prop-qto-1} ensures that \AnC{} holds so that
we may analytically continue the left-hand side of this identity and conclude that, for
$\alpha\in\fS(\vartheta)$,
$$
[D\omega_t:D\omega]_\alpha^{-1}=[D\omega_t:D\omega]_{-\bar\alpha}^\ast.
$$
In particular, since $\vartheta>1/2$ by assumption, we have
$$
B:=[D\omega_T:D\omega]_{\frac12}\in\cO,\quad\text{and}\quad
B^{-1}=[D\omega_T:D\omega]^\ast_{-\frac12}.
$$
Further, the estimate~\eqref{est-sa} yields
\beq
D_T=\e^{-2|T|(\|\sigma_\omega^{\i/2}(V)\|+\|V\|)}\leq
\frac1{\|B^{-1}\|^2}\le B^\ast B\le
 \|B\|^2\leq \e^{2|T|(\|\sigma_\omega^{-\i/2}(V)+\|V\|)}=C_T.
\label{eq:Bbound}
\eeq

The identities
\[
\Delta_{\omega_T|\omega}^{1/2}\Omega=\Omega_{\omega_T},
\]
\[
[D\omega_T:D\omega]_\alpha\Omega=\Delta_{\omega_T|\omega}^{\alpha}\Omega,
\qquad \alpha \in \i\RR,
\]
and analytic continuation give that
\[
B\Omega=\Omega_{\omega_T},
\]
and hence
\[
\Omega_{\omega_T}=J\Omega_{\omega_T}=JB\Omega=JBJ\Omega=B'\Omega
\]
where $B'=JBJ\in\fM'$. For $\alpha\in]-\vartheta,\vartheta[$ and $\theta\in\RR$, we set
\[
A(\alpha,\theta):=\varsigma_\omega^\theta\left(
[D{\omega_{t}}:D{\omega}]_{\frac{\bar \alpha}2}^\ast
[D{\omega_{t}}:D{\omega}]_{\frac\alpha2}\right).
\]
Noticing that $A(\alpha,\theta)$ is a self-adjoint and strictly positive element
of $\cO$, we derive
$$
\omega_T\circ\varsigma_\omega^\theta\left([D{\omega_{t}}:D{\omega}]_{\frac{\bar \alpha}{2} }^\ast
[D{\omega_{t}}:D{\omega}]_{\frac{\alpha}2}\right)
=\langle B'\Omega,A(\alpha,\theta) B'\Omega\rangle
=\langle A(\alpha,\theta)^{\frac12}\Omega, JB^\ast BJA(\alpha,\theta)^{\frac12}\Omega\rangle
$$
and hence, invoking the inequalities~\eqref{eq:Bbound},
\begin{align*}
D_T \omega\left([D{\omega_{t}}:D{\omega}]_{\frac{\bar \alpha}2}^\ast
[D{\omega_{t}}:D{\omega}]_{\frac\alpha2}\right)
 & \le \omega_T\circ\varsigma_\omega^\theta\left([D{\omega_{t}}:D{\omega}]_{\frac{\bar \alpha}{2} }^\ast
[D{\omega_{t}}:D{\omega}]_{\frac{\alpha}2}\right) \\[2mm]
 &  \le C_T \omega\left([D{\omega_{t}}:D{\omega}]_{\frac{\bar \alpha}2}^\ast
[D{\omega_{t}}:D{\omega}]_{\frac\alpha2}\right).
\end{align*}
Setting $\theta=0$ yields the inequalities~\eqref{proja-2}. Invoking Formula~\eqref{toul-late},
we get the inequalities~\eqref{t-na-1} by averaging over $\theta$.

\subsection{Proof of Proposition~\ref{start-qpsc-1}}
\label{sec-proof-start-qpsc-1}

By the chain rule~\eqref{eq:chain}, we have
$$
[D\omega_{t+s}:D\omega]_{\alpha}=[D\omega_{t+s}:D\omega_t]_\alpha[D\omega_t:D\omega]_{\alpha}
$$
for any $s,t\in\RR$ and $\alpha\in\i\RR$. Invoking
Definition~\eqref{eq:ConnesDef} and the covariance
relation~\eqref{eq:DeltaCovar}, we further derive
\begin{align*}
[D\omega_{t+s}:D\omega_t]_\alpha
& =\Delta_{\omega_s\circ\tau^t|\omega\circ\tau^t}^\alpha\Delta_{\omega\circ\tau^t}^{-\alpha}\\[2mm]
& =\e^{-\i t\cL}\Delta_{\omega_s|\omega}^\alpha\Delta_{\omega}^{-\alpha}\e^{\i t\cL}\\[2mm]
& =\e^{-\i t\cL}[D\omega_s:D\omega]_\alpha\e^{\i t\cL}\\[2mm]
& =\tau^{-t}([D\omega_s:D\omega]_\alpha),
\end{align*}
from what the result follows.

\subsection{Proof of Theorem~\ref{thm-qspc-ti-2}}
\label{sec-proof-thm-qpsc}

(1) follows from~\QuOne{} by differentiating the cocycle relation~\eqref{tom-vac-1} at $\alpha=0$.

\noindent(2) Let $\Phi\in\Dom(\log\Delta_{\omega_t|\omega})$ and $\Psi\in\Dom(\log\Delta_\omega)$.
Differentiating the identity
\[
\langle\Phi,[D\omega_t:D\omega]_{\i\theta}\Psi\rangle
=\langle\e^{-\i\theta\log\Delta_{\omega_t|\omega}}\Phi,\e^{-\i\theta\log\Delta_\omega}\Psi\rangle
\]
w.r.t.\;$\theta$ at $\theta=0$ gives
\[
\langle \log \Delta_{\omega_t|\omega}\Phi,\Psi\rangle
=\langle \Phi, (\log \Delta_\omega +\ell_{\omega_t|\omega})\Psi\rangle,
\]
and the first assertion follows. To prove the second one we observe that,
starting with~\eqref{eq:RelEntDef} and invoking the covariance
relation~\eqref{eq:DeltaCovar}, we get
$$
\Ent(\omega_t|\omega)=\langle\Omega_{\omega_t},\log\Delta_{\omega|\omega_t}\Omega_{\omega_t}\rangle
=\langle\e^{-\i t\cL}\Omega,\log\Delta_{\omega|\omega_t}\e^{-\i t\cL}\Omega\rangle
=\langle\Omega,\log\Delta_{\omega_{-t}|\omega}\Omega\rangle.
$$
By the first assertion and Definition~\eqref{eq:ctDef}, we further get
$$
\Ent(\omega_t|\omega)=\langle\Omega,(\log\Delta_\omega+\ell_{\omega_{-t}|\omega})\Omega\rangle
=\langle\Omega,\tau^t(c^{-t})\Omega\rangle.
$$
Finally, \eqref{str-0-quantum} gives that $\tau^t(c^{-t})=c^t$, and the second
assertion follows.

\noindent(3) Assumption~\QuTwo{} and Part~(1) give
\beq
\frac{\d\ }{\d t}c^t=\frac{\d\ }{\d s}c^{t+s}\bigg|_{s=0}
=\frac{\d\ }{\d s}\left(c^t+\tau^t(c^s)\right)\bigg|_{s=0}
=\tau^t(\sigma),
\label{eq:proofthm2.18}
\eeq
from which the first assertion immediately follows. Identity
\eqref{eq:proofthm2.18} combined with the second assertion of Part~(2) give the
entropy balance equation~\eqref{ent-balance-2-1}.

\noindent(4) We have
\beq
1=\langle\Omega_{\omega_t},\Omega_{\omega_t}\rangle
=\langle\Delta_{\omega_t|\omega}^{1/2}\Omega,\Delta_{\omega_t|\omega}^{1/2}\Omega\rangle
=\langle\e^{(\log\Delta_\omega+\ell_{\omega_t|\omega})/2}\Omega,
\e^{(\log\Delta_\omega+\ell_{\omega_t|\omega})/2}\Omega\rangle.
\label{vac-diff}
\eeq
Araki's perturbation theory of the KMS-structure gives
\[
\frac{\d\ }{\d t}\e^{(\log\Delta_\omega+\ell_{\omega_t|\omega})/2}\Omega\bigg|_{t=0}
=\int_0^{1/2}\Delta_\omega^s\sigma\Omega\d s,
\]
see {\sl e.g.}~\cite[Relation~(7.3)]{Jaksic2014b}.
Differentiating~\eqref{vac-diff} w.r.t.\;$t$ at $t=0$ thus yields
$$
0=2\Re\int_0^{1/2}\langle\Delta_\omega\Omega,\Delta_\omega^s\sigma\Omega\rangle\d s
=\omega(\sigma).
$$

\noindent(5) By~\cite[Proposition~2.2]{Benoist2023a}, there exists an
anti-unitary operator $W$ on the GNS Hilbert space $\cH$ such that
$\Theta(A)=WAW^\ast$ for all $A\in\cO$ and
$W\Delta_{\omega_t|\omega}W^\ast=\Delta_{\omega_{-t}|\omega}$ for all $t\in\RR$.
It follows that for $\alpha\in\i\RR$ and $t\in\RR$,
\[
\Theta\left([D\omega_{t}:D\omega]_{\alpha}\right)=[D\omega_{-t}:D\omega]_{-\alpha}.
\]
Differentiating this identity at $\alpha=0$ gives
$\Theta(\ell_{\omega_t|\omega})=\ell_{\omega_{-t}|\omega}$ and hence
$$
\Theta(c^t)
=\Theta\circ\tau^t(\ell_{\omega_t|\omega})
=\tau^{-t}\circ\Theta(\ell_{\omega_t|\omega})
=\tau^{-t}(\ell_{\omega_{-t}|\omega})=c^{-t}.
$$
Differentiating now at $t=0$ gives $\Theta(\sigma)=-\sigma$.

\subsection{Proof of Proposition~\ref{prop-qto-2}}
\label{sec-proof-qto-2}

\noindent(1) First we note that~\AnC{} and analytic continuation give
that~\eqref{tom-vac-1} holds for all $\alpha\in\fS(\vartheta)$. We then have
\begin{align*}
U_\alpha^{t+s}&=\e^{\i(t+s)\cL}J[D\omega_{t+s}:D\omega]_{\bar\alpha}J\\[2mm]
&=\e^{\i(t+s)\cL}J\e^{-\i s\cL}[D\omega_t:D\omega]_{\bar\alpha}\e^{\i s\cL}[D\omega_s:D\omega]_{\bar \alpha}J\\[2mm]
&=\left(\e^{\i t\cL}J[D\omega_t:D\omega]_{\bar\alpha}J\right)\left(\e^{\i s\cL}J[D\omega_s:D\omega]_{\bar\alpha}J\right)\\[2mm]
&=U_\alpha^tU_\alpha^s,
\end{align*}
where we used that, for any $s\in\RR$, $J\e^{-\i s \cL}=\e^{-\i s \cL}J$. This
yields the group property. To prove that this group has the $C_0$-property,
invoking~\cite[Proposition 1.18]{Davies1980} it suffices to show that, for all
$\Psi\in\cH$,
\beq
\lim_{t\to0}U_\alpha^t\Psi=\Psi.
\label{mar-hun-1}
\eeq
By Vitali's convergence theorem, combined with~\AnC{} and the
bound~\eqref{mar-hun}, it is sufficient to prove~\eqref{mar-hun-1} for
$\alpha\in\i\RR$. Using the definition of the Araki--Connes cocycle we can write
$$
U_\alpha^t-I
=\e^{\i t\cL}J\Delta_{\omega_t|\omega}^{-\alpha}\Delta_\omega^\alpha J-I
=\e^{\i t\cL}J\Delta_{\omega_t|\omega}^{-\alpha}(\Delta_\omega^\alpha
  -\Delta_{\omega_t|\omega}^{\alpha})J+(\e^{\i t\cL}-I),
$$
which leads to the bound
$$
\|(U_\alpha^t-I)\Psi\|
\le\|(\Delta_{\omega_t|\omega}^\alpha-\Delta_\omega^\alpha)J\Psi\|
+\|(\e^{\i t\cL}-I)\Psi\|,
$$
so that it suffices to show that
\beq
\slim_{t\to0}\Delta_{\omega_t|\omega}^\alpha=\Delta_\omega^\alpha.
\label{eq:satfoo}
\eeq
By a well known inequality~\cite[Theorem~2.5.31(b)]{Bratteli1987},
$\|\omega_t-\omega\|\le2\|(\e^{-\i t\cL}-I)\Omega\|\to0$ as $t\to0$, and it
follows from~\cite[Lemma~4.1]{Araki1977} that
$\Delta_{\omega_t|\omega}^{1/2}\to\Delta_\omega^{1/2}$ in the strong resolvent
sense. Applying the well known result~\cite[Theorem~VIII.20(b)]{Reed1980} to the
bounded continuous function $[0,\infty[\ni x\mapsto x^{\i t}$ yields~\eqref{eq:satfoo},
thus establishing~\eqref{mar-hun-1}.

\noindent(2) For $\alpha\in\i\RR$ the operator $U_{\alpha}^t$ is the product of
two unitaries, and hence is unitary.

\noindent(3) By analyticity, it suffices to prove the statement for
$\alpha\in\i\RR$. Then
\[
U_\alpha^tAU_{-\bar\alpha}^{t\ast}
=U_\alpha^tAU_\alpha^{t\ast}
=\e^{\i t\cL}J[D\omega_t:D\omega]_{\bar\alpha}JAJ[D\omega_t:D\omega]^\ast_{\bar\alpha}J\e^{-\i t\cL},
\]
and the statement then follows from the fact that $JAJ\in\fM'$ commutes with the
unitary $[D\omega_t:D\omega]_{\bar\alpha}\in\fM$.

\noindent(4) Again, by analyticity it suffices to prove the statement for $\alpha\in\i\RR$
in which case $-\bar\alpha=\alpha$. The identity~\eqref{tom-vac-1} gives
\[
\one=\tau^{-t}([D\omega_{-t}:D\omega]_{\bar \alpha})[D\omega_{-t}:D\omega]_{\bar \alpha},
\]
and so
\beq
[D\omega_t:D\omega]_{\bar\alpha}^\ast\e^{-\i t\cL}
=\e^{-\i t\cL}[D\omega_{-t}:D\omega]_{\bar \alpha}.
\label{ns-hu}
\eeq
Since $J\e^{-\i t\cL}=\e^{-\i t\cL}J$,  Relation~\eqref{ns-hu} gives
\[
J[D\omega_t:D\omega]_{\bar\alpha}^\ast J\e^{-\i t\cL}
=\e^{-\i t\cL}J[D\omega_t:D\omega]_{\bar \alpha}J,
\]
and the statement follows.

\noindent(5) The first identity in~\eqref{sun-ho} follows from Relation~\eqref{an-sa}.
The second one follows from the first and Relation~\eqref{eq:StdLiouvPert}.

\subsection{Proof of Proposition~\ref{prop-qto-4}}
\label{sec-proof-qto-1}

\noindent(1) We will prove a stronger statement, namely that
\beq
\e^{\i t\cL_{\shalf-\alpha}}\Omega=[D\omega_{-t}:D\omega]_\alpha\Omega.
\label{hhis-1}
\eeq
Recalling that $\Delta_{\omega_t|\omega}^{1/2}\Omega=\Omega_{\omega_t}$,
the definitions of the transfer operator and the Araki--Connes cocycle give
\begin{align*}
\e^{\i t\cL_{\shalf-\alpha}}\Omega
=U_{\shalf-\alpha}^t\Omega
&=\e^{\i t\cL}J[D\omega_t:D\omega]_{\frac12-\bar\alpha}J\Omega\\[4pt]
&=\e^{\i t\cL}J\Delta_{\omega_t|\omega}^{1/2-\bar\alpha}\Delta_\omega^{-1/2+\bar\alpha}J\Omega
=\e^{\i t\cL}J\Delta_{\omega_t|\omega}^{1/2-\bar\alpha}\Omega
=\e^{\i t\cL}J\Delta_{\omega_t|\omega}^{-\bar\alpha}\Omega_{\omega_t}.
\end{align*}
Using the fact that $\Omega_{\omega_t}=\e^{-\i t\cL}\Omega$ and
Relation~\eqref{eq:DeltaCovar}, we get
\begin{align*}
\e^{\i t\cL_{\shalf-\alpha}}\Omega
& =\e^{\i t\cL}J\Delta_{\omega_t|\omega}^{-\bar\alpha}\e^{-\i t\cL}\Omega
=J\e^{\i t\cL}\Delta_{\omega_t|\omega}^{-\bar\alpha}\e^{-\i t\cL}\Omega \\[2mm]
& =J\Delta_{\omega|\omega_{-t}}^{-\bar\alpha}J\Omega
=\Delta_{\omega_{-t}|\omega}^\alpha\Omega
=[D\omega_{-t}:D\omega]_\alpha\Omega,
\end{align*}
and hence
$$
\fF_{\omega,t}^\ttm(\alpha)=\omega([D\omega_{-t}:D\omega]_\alpha)
=\langle\Omega,[D\omega_{-t}:D\omega]_\alpha\Omega\rangle
=\langle\Omega,\e^{\i t \cL_{\shalf-\alpha}}\Omega\rangle.
$$

\noindent(2) Relations~\eqref{hhis} and~\eqref{hhis-1} yield
\[
\fF_{\omega_T,t}^\qpsc(\alpha)
=\omega_T([D\omega_{-t}:D\omega]_\alpha)
=\langle\Omega,\e^{\i T\cL_{\shalf}}[D\omega_{-t}:D\omega]_\alpha\Omega\rangle
=\langle\Omega,\e^{\i T\cL_{\shalf}}\e^{\i t\cL_{\shalf-\alpha}}\Omega\rangle.
\]

\noindent(3) It follows from~\eqref{hhis-1} that
\beq
\Delta_\omega^{-\alpha/2}\e^{\i t\cL_{\shalf-\alpha}}\Omega
=\Delta_\omega^{-\alpha/2}\Delta_{\omega_{-t}|\omega}^\alpha\Delta_\omega^{-\alpha/2}\Omega=
[D\omega_{-t}: D\omega]^\ast_{\frac{\bar \alpha}{2}}[D\omega_{-t}: D\omega]_{\frac{\alpha}{2}}\Omega,
\label{wiwb}
\eeq
where we used that $\bar \alpha =-\alpha$ for $\alpha \in \i \RR$. Relations~\eqref{hhis} and~\eqref{wiwb} further yield
\begin{align*}
\fF_{\omega_T,t}^\anc(\alpha)
&=\omega_T\left([D\omega_{-t}:D\omega]^\ast_{\frac{\bar \alpha}{2}}[D\omega_{-t}:D\omega]_{\frac{\alpha}{2}}\right)\\[4pt]
&=\langle\Omega,\e^{\i T\cL_\shalf}\Delta_\omega^{-\alpha/2}\e^{\i t\cL_{\shalf-\alpha}}\Omega\rangle
=\langle\Omega,\e^{\i T\cL_\shalf}\Delta_\omega^{-\alpha/2}\e^{\i t\cL_{\shalf-\alpha}}\Delta_\omega^{\alpha/2}\Omega\rangle,
\end{align*}
and the desired identity follows immediately from Relation~\eqref{sun-tuluz}.

\noindent(4) The stated analytic extensions all follow from Assumption~\AnV{} and
 Definitions~\eqref{sn-mon-1}, \eqref{qpscDef}, \eqref{sun-ho} and~\eqref{eq:hatLDef}.

\subsection{Proof of Proposition~\ref{first-gen-sem-1}}
\label{sec-int-1}

The argument follows the proof of Paley--Wiener's theorem~\cite[Theorem 19.2]{Rudin1987}.
We provide the details.

Recall that $m,M$ are given by~\eqref{timed}. For $R\geq R_0=|\fr|+r+1$ and $a>\gamma+m$, let
$\Gamma_R$ be the rectangle with vertices $\{\pm R-\i\gamma,\pm R+\i a\}$, and note that
$\fr$ is inside $\Gamma_R$. The only singularity of the function
\[
z\mapsto f(z)=\bra\phi,(z-L)^{-1}\psi\ket
\]
inside $\Gamma_R$ is a pole of order $N\ge1$ at $\fr$ and so
$$
\frac1{2\pi i}\oint_{\Gamma_R}\e^{-\i tz}f(z)\d z
=p(t)\e^{-\i t\fr},
$$
where $p$ is a polynomial of degree $N-1$ such that $p(0)$ is the residue of $f$ at $\fr$.
Note that, in particular, $p$ is constant and equals to this residue if $N=1$.
We analyze separately the integrals over the four sides of the rectangle $\Gamma_R$
as $R\to\infty$ along a well-chosen sequence, assuming that $t\geq 1$.

Consider first the integral over the bottom side of $\Gamma_R$.
An integration by parts gives
\beq
I_1(R,t)
=\int_{-R}^R\e^{-\i t(x-\i\gamma)}f(x-\i\gamma)\d x
=\frac{\e^{-\gamma t}}{\i t}\left[\int_{-R}^R\e^{-\i tx}f'(x-\i\gamma)\d x+B\right]
\label{so-l}
\eeq
where
\[
B=\e^{\i Rt}f(-R-\i\gamma)-\e^{-\i Rt}f(R-\i\gamma).
\]
Assumption~\eqref{eq:resolventestimate} for $j=1$ gives that both the integral and
the boundary terms $B$ in~\eqref{so-l} are uniformly bounded with respect to
$R\geq R_0$ and $t\geq 1$. More precisely, one has the bound
\beq
\sup_{R\geq R_0}|I_1(R,t)|\leq K\e^{-\gamma t},
\label{mar-r}
\eeq
where
\beq
K=2\int\limits_{|x|>r}|f'(x-\i\gamma)|\d x+\int\limits_{|x|<r}|f'(x-\i\gamma)|\d x
+|f(-R_0-\i \gamma)|+|f(R_0-\i\gamma)|.
\label{K-imp}
\eeq

We now consider the integrals over the vertical sides $\mp R+\i[-\gamma,a]$. Let
\[
I_2(\pm R,t)=\pm\int_{-\gamma}^a\e^{\i t(\pm R+\i y)}f(\pm R+\i y)\d y.
\]
The Cauchy-Schwarz inequality gives
\[
|I_2(\pm R,t)|^2\leq\left(\int_{-\gamma}^a \e^{-2ty}\, \d y\right) g(\pm R),
\]
where
\[
g(R)=\int_{-\gamma}^a |f(\pm R+\i y)|^2\d y.
\]
Assumption~\eqref{eq:resolventestimate} for $j=0$ yields that the function $R\mapsto g(-R)+g(R)$
is integrable on $[R_0,+\infty[$, and so there exists a sequence $(R_k)_{k\in\NN^\ast}$
such that $\lim_{k\to\infty}R_k=\infty$ and $\lim_{k\to\infty} g(\pm R_k)=0$. It follows that
\begin{equation}
\lim_{k\to\infty}I_2(\pm R_k,t)=0.
\label{mar-r-1}
\end{equation}

We now consider the integral over the top side of $\Gamma_R$,
\[
I_3(R,t)=-\e^{at}\int_{-R}^R\e^{-\i tx}f(x+\i a)\d x.
\]
Since $a >m$ we have that, for $x\in\RR$,
\[
f(x+\i a)
=\bra \phi,(x+\i a-L)^{-1}\psi\ket = -\i\int_0^{\infty} \e^{\i t (x+\i a)} \bra \phi, \e^{- \i t L}\psi\ket \, \d t,
\]
and so the left-hand side is the Fourier transform of the $L^2$-function
\[
t\mapsto-\i\e^{-at}\bra\phi,\e^{-\i tL}\psi\ket\one_{]0,\infty[}(t).
\]
The Plancherel theorem gives that
\[
\lim_{k\to\infty}I_3(R_k,t)
=2\pi\i\bra\phi,\e^{-\i tL}\psi\ket\one_{]0,\infty[}(t)
\]
where the limit is in $L^2(\RR,\d t)$ and the $R_k$ are as in~\eqref{mar-r-1}.
Hence, there exists a subsequence $(R_{k_n})_{n\in\NN}$ such that for a.e.\;$t\geq 1$
$$
\lim_{n\to\infty}I_3(R_{k_n},t)=2\pi\i\bra\phi,\e^{-\i t L}\psi\ket.
$$
Combining~\eqref{mar-r} and~\eqref{mar-r-1} we derive that for a.e.\;$t\geq 1$, and with $K$ given by~\eqref{K-imp},
\beq
|\bra\phi,\e^{-\i tL}\psi\ket-\e^{-\i\fr t}p(t)
|\leq K\e^{-\gamma t}.
\label{final-s}
\eeq
Since both sides in~\eqref{final-s} are continuous w.r.t.\;$t$, \eqref{final-s} holds for
all $t\geq 1$.

\subsection{Proof of Proposition~\ref{second-gem-sem-1}}
\label{sec-int-2}

Since $\bra\phi,\e^{-\i tL}\psi\ket=\e^{-\i t\fr}p(t)+R(t)$, with $p$ a polynomial of degree $N-1$ and
\beq
|R(t)|\le K\e^{-\gamma t},
\label{eq:Testim}
\eeq
we have, for $z\in\CC$ such that $\Im z>m$,
\beq
\bra\phi,(z-L)^{-1}\psi\ket
=-\i\int_0^\infty\e^{\i tz}\bra\phi,\e^{-\i tL}\psi\ket\d t
=\frac{q(z-\fr)}{(z-\fr)^N}-\i\int_0^\infty\e^{\i zt}R(t)\d t,
\label{so-h}
\eeq
where $q$ is a polynomial of degree $N-1$. From the estimate~\eqref{eq:Testim},
we deduce that the Laplace transform
\[
z\mapsto\int_0^\infty\e^{\i zt}R(t)\d t
\]
has an analytic continuation to the half-plane $\Im z>-\gamma$. This proves the first
part of the proposition.

Turning to the resolvent estimates~\eqref{eq:resolventestimate2}, we deal first with the case
$j=0$. Let $0<\mu<\gamma$ and $r>|\fr|+1$. It follows from~\eqref{so-h} that for all
$y>-\mu$ and $|x|>r$ we can write
\beq
\bra\phi,(x+\i y-L)^{-1}\psi\ket=\frac{q(x-\fr+\i y)}{(x-\fr+\i y)^N}+\hat g(x),
\label{eq:gEq}
\eeq
where
\[
\hat g(x)=-\i\int_0^\infty\e^{\i tx}\e^{-ty}R(t)\d t.
\]
There is a constant $C>0$ such that, in the region $|x|>r$,
the modulus of the first term on the right-hand side
of Relation~\eqref{eq:gEq} is bounded, uniformly in $y$, by the function
$x\mapsto\frac{C}{|x-\Re\fr|}$, which is square integrable on this region.
The second term is the Fourier transform of the function
\[
g(t)=-\i\e^{-ty}R(t)\one_{]0,\infty[}(t),
\]
which is such that
$$
\int_\RR|g(t)|^2\d t\le\frac{K^2}{2(\gamma+y)}\le\frac{K^2}{2(\gamma-\mu)}.
$$
The Plancherel theorem gives that the $L^2$-norm of $\hat g$ is uniformly bounded for $y>-\mu$.

In the case $j=1$ we note that, for some polynomial $\tilde q$ of degree $N-1$,
\[
\partial_x\bra\phi,(x+\i y-L)^{-1}\psi\ket
=\frac{\tilde q(x-\fr+\i y)}{(x-\fr+\i y)^{N+1}}+\int_0^\infty\e^{\i tx}\e^{-ty}tR(t)\d t,
\]
where the first term on the right-hand side is bounded, uniformly in $y$, by a function
$x\mapsto\frac{\tilde C}{|x-\Re\fr|^2}$ which is integrable on the region $|x|>r$. To deal with
the second term we perform two integrations by parts, obtaining
$$
\int_0^\infty\e^{\i tx}\e^{-ty}tR(t)\d t
=-\frac1{(x+\i y)^2}\left(R(0)
+\int_0^\infty\e^{\i tx}\e^{-ty}(2R^{(1)}(t)+tR^{(2)}(t))\d t
\right),
$$
which is clearly integrable on the region $|x|>r$, uniformly for $y>-\mu$.

\subsection{Proof of Theorem~\ref{main-spect-def}}

By Proposition~\ref{prop-qto-4}(1+4), the formula
\[
\fF_{\omega,t}^\ttm(\alpha)=\langle\Omega,\e^{\i t\cL_{\shalf-\alpha}}\Omega\rangle
\]
holds for $\alpha$ satisfying $\left|\Re\left(\alpha-\frac12\right)\right|<\vartheta$. Assumption~\DeformTwo(d)
and Proposition~\ref{jpr-1} applied to $\phi=\psi=\Omega$ and $L=-\cL_{\shalf-\alpha}$ give that
$$
\langle\Omega,\e^{\i t\cL_{\shalf-\alpha}}\Omega\rangle=\e^{-\i t\cE(\frac12-\alpha)}p_\alpha(t)+O({\e^{-t\gamma}}),
$$
where $p_\alpha$ is a non-zero polynomial by the first condition in~\DeformTwo(g)\footnote{Recall
from the proof of Proposition~\ref{jpr-1} that $p_\alpha(0)=\langle\Omega,\cR_{\frac12-\alpha}\Omega\rangle$.}
and $\gamma>-\Im\cE(\frac12-\alpha)$. It follows that
\beq
\lim_{t\to\infty}\tfrac1t\log\langle\Omega,\e^{\i t\cL_{\shalf-\alpha}}\Omega\rangle
=-\i\cE\left(\tfrac12-\alpha\right)
\label{i-can}
\eeq
for $-\vartheta+\frac12<\Re\alpha<\vartheta+\frac12$ and $|\Im\alpha|<\zeta$. This gives the existence of the first limit
in~\eqref{thup-thup-1}. Part~(1) follows from~\eqref{i-can} and the observation that
$\fF_{\omega,t}^\ttm(\alpha)>0$ for $\alpha\in]-\vartheta,\vartheta[$.

Next, Definition~\eqref{qpscDef}, Theorem~\ref{thm-ness} and~\DeformTwo(a) give that
$$
\fF^\qpsc_{\omega_+,t}(\alpha)
=\omega_+([D\omega_{-t}:D\omega]_{\alpha})
=\langle\Omega,\cR_\shalf D^{-1}[D\omega_{-t}:D\omega]_\alpha\Omega\rangle,
$$
for $\alpha\in B(\vartheta,\zeta)$. Invoking Relation~\eqref{hhis-1} we further obtain
$$
\fF^\qpsc_{\omega_+,t}(\alpha)
=\langle D^{-1}\cR_\shalf^\ast\Omega,\e^{\i t\cL_{\shalf-\alpha}}\Omega\rangle.
$$
Assumption~\DeformTwo(c) allows us to write
\[
\langle D^{-1}\cR_\shalf^\ast\Omega,(z+\cL_{\shalf-\alpha})^{-1}\Omega\rangle
=\langle D^{-2}\cR_\shalf^\ast\Omega,D(z+\cL_{\shalf-\alpha})^{-1}D\Omega\rangle,
\]
and so we can combine Assumption~\DeformTwo(d+e+g) and Proposition~\ref{jpr-1} with
$\phi=D^{-1}\cR_\shalf^\ast\Omega$, $\psi=\Omega$, and $L=-\cL_{\shalf-\alpha}$.
This yields the third limit in~\eqref{thup-thup-1}
for $-\vartheta+\frac12<\Re\alpha<\vartheta$ and $|\Im\alpha|<\zeta$.

The second limit is handled by a very similar argument.
Definition~\eqref{sn-mon-1}, Theorem~\ref{thm-ness} and~\DeformTwo(b) give that,
for $\alpha\in B(\vartheta,\zeta)$,
$$
\fF^\anc_{\omega_+,t}(\alpha)
=\omega_+([D\omega_{-t}:D\omega]^\ast_{\frac{\bar\alpha}2}[D\omega_{-t}:D\omega]_{\frac\alpha2})
=\langle D^{-1}\cR_\shalf^\ast\Omega,[D\omega_{-t}:D\omega]^\ast_{\frac{\bar\alpha}2}
[D\omega_{-t}:D\Omega]_{\frac\alpha2}\Omega\rangle.
$$
Relation~\eqref{wiwb} further gives
$$
\fF^\anc_{\omega_+,t}(\alpha)
=\langle\Delta^{-\bar\alpha/2}D^{-1}\cR_\shalf^\ast\Omega,\e^{\i t\cL_{\shalf-\alpha}}\Omega\rangle.
$$
Again~\DeformTwo(c) allows us to write
\[
\langle\Delta^{-\bar\alpha/2}D^{-1}\cR_\shalf^\ast\Omega,(z+\cL_{\shalf-\alpha})^{-1}\Omega\rangle
=\langle D^{-1}\Delta^{-\bar\alpha/2}D^{-1}\cR_\shalf^\ast\Omega,D(z+\cL_{\shalf-\alpha})^{-1}D\Omega\rangle,
\]
and, invoking~\DeformTwo(f), we conclude the proof in the same way as in the case of the third limit.

\subsection{Proof of Theorem~\ref{ss-tuluz}}

It suffices to show that, given $\alpha_0\in B(\vartheta,\zeta)$ and a small enough open
neighborhood $U\ni\alpha_0$ satisfying Conditions~\DeformThree, the function
$\alpha\mapsto\cE(\alpha)$ is analytic in $U$. We set $R=\sup_{\alpha\in U}|\cE(\alpha)|+r_U+1$
and fix $\gamma$ such that
$$
\max(0,-\inf_{\alpha\in U}\Im\cE(\alpha))<\gamma<\mu.
$$
By Assumptions~\DeformThree(b+c) one has
$$
K=\sup_{\alpha\in U}\left(2\int_{-\infty}^{+\infty}|\langle\Omega,(x-\i\gamma+\cL_\alpha)^{-2}\Omega\rangle|\d x
+|\langle\Omega,(-R-\i\gamma+\cL_\alpha)^{-1}\Omega\rangle|
+|\langle\Omega,(+R-\i\gamma+\cL_\alpha)^{-1}\Omega\rangle|\right)<\infty.
$$
We first note that $\alpha\mapsto f_t(\alpha)=\langle\Omega,\e^{\i t\cL_\alpha}\Omega\rangle$
is analytic by~\eqref{tpar-L} and~\AnC. From the proof of Proposition~\ref{first-gen-sem-1}
we further get that
$$
f_t(\alpha)=\e^{-\i t\cE(\alpha)}p_\alpha(t)+F_t(\alpha),
$$
where $p_\alpha$ is a non-zero polynomial and
$\sup_{\alpha\in U}|F_t(\alpha)|\le K\e^{-\gamma t}$. Thus, for large enough $T>0$, one has
$\left\{f_t(\alpha)\mid t\ge T,\alpha\in U\right\}\subset\CC^\ast$. It follows that, for $t\ge T$, $\log f_t(\alpha)$
has an holomorphic branch satisfying
$$
\tfrac1t\log f_t(\alpha)=-\i\cE(\alpha)+O(t^{-1}),
$$
so that Weierstrass convergence theorem allows us to conclude that $\cE$ is holomorphic on $U$.

\printbibliography[heading=bibintoc,title={References}]
\end{document}